\documentclass[useAMS,usenatbib]{mn2e}
\usepackage{graphicx}
\usepackage{amsmath}
\usepackage{textcomp}
\newcommand{\hb}{H$\beta$ }

\newcommand{\ot}{[OIII] }
\newcommand{\kms}{{\, \rm km~s}^{-1}}

\title[Study of a homogeneous QSO sample: relations between the QSO and its host galaxy]{Study of a homogeneous QSO sample: relations between the QSO and its host galaxy\thanks{Based on observations collected at the European Organisation for Astronomical Research in the Southern Hemisphere, Chile, under program IDs 77.B-0229, and 78.B-0081.}}
\author[Y. Letawe]{Y. Letawe$^{1}$%\footnotemark[1]\thanks{Thank you so much}
, G. Letawe$^{1}$\thanks{E-mail:
gletawe@ulg.ac.be} and P. Magain$^{1}$\\%\footnotemark[1]\thanks{This file has been amended to highlight the proper use of \LaTeXe\ code with the class file. These changes are for illustrative purposes and do not reflect the original paper by A. V. Raveendran.}\\
$^{1}$D\'epartement d'Astrophysique, G\'eophysique et Oc\'eanographie, Ulg, All\'ee du 6 ao\^{u}t, 13, 4000 Sart Tilman, Belgium}
\begin{document}

\date{Accepted 2009 December 15. Received 2009 December 14; in original form 2009 October 11}

\pagerange{\pageref{firstpage}--\pageref{lastpage}} \pubyear{2009}

\maketitle

\label{firstpage}

\begin{abstract}
We analyse a sample of $69$ QSOs which have been randomly selected in a complete sample of $104$ QSOs ($R\leq18$, $0.142<z<0.198$, $\delta < 10^{o}$). $60$ have been observed with the NTT/SUSI2 at La Silla, through two filters in the optical band (WB$\#655$ and V$\#812$), and the remaining $9$ are taken from archive databases. The filter V$\#812$ contains the redshifted $H\beta$ and forbidden \ot emission lines, while WB$\#655$ covers a spectral region devoid of emission lines, thus measuring the QSO and stellar continua. 
The contributions of the QSO and the host are separated thanks to the MCS deconvolution algorithm, allowing a morphological classification of the host, and the computation of several parameters such as the host and nucleus absolute V-magnitude, distance between the luminosity center of the host and the QSO, and colour of the host and nucleus. We define a new asymmetry coefficient, independent of any galaxy models and well suited for QSO host studies. The main results from this study are: (i) $25\%$ of the total number of QSO hosts are spirals, $51\%$ are ellipticals and $60\%$ show signs of interaction; (ii) Highly asymmetric systems tend to have a higher gas ionization level (iii) Elliptical hosts contain a substantial amount of ionized gas, and some show off-nuclear activity. These results agree with hierarchical models merger driven evolution.  
%Moreover, we show that: (i) The asymmetry coefficient is a good indicator of the degree of interaction;
%\begin{enumerate}
%\item $M_V(Host)$ slightly correlates with $M_V(QSO)$, in agreement with previous results;
%\item The asymmetry coefficient is a good indicator of the degree of interaction;
%\item Highly asymmetric systems tend to have a higher gas ionization level;
%\item $25\%$ of the total number of QSO hosts are spirals, $51\%$ are ellipticals and $60\%$ show signs of interaction; 
%\item $4$ systems contain more than one nucleus;
%\item $12$ systems ($21\%$) are too distorted or asymmetric to be identified as spirals or ellipticals, inviting for further studies with higher resolution apertures;
%\item Elliptical hosts contain a substantial amount of ionized gas, and some show off-nuclear activity. This agrees with hierarchical models merger driven evolution.
%\end{enumerate}

\end{abstract}


\section{Introduction}

Over the past decade, many observations and studies have been put forwards in order to reach a better understanding of the interrelations between the QSOs and their hosts. This has been made possible thanks to the availability of high resolution space based data, as well as ground based imaging. Several strong correlations have been found, such as the famous black hole mass - bulge stellar velocity dispersion relation for quiescent galaxies \citep{Ferrarese,b2,bernardi}, extended afterwards to active galaxies \citep{McLure2001,McLure2002,Marconi}. The most popular global picture that has emerged is to place the QSO phenomenon in an evolutionary context, in which galaxies evolve hierarchically by successive gravitational interactions or mergers, allowing more gas to reach the central regions and possibly trigger the QSO phase. Even if the observational basis of this idea remains unsecure, convincing hints of merger induced QSO activity were given using either hydrodynamical simulations \citep{Hopkins2005,Hopkins2006}, high resolution imaging \citep{Bahcall97,Bennert,LetaweY}, or 2D-spectroscopy \citep{Letawe}. On the other hand, \citet{Schmitt} argued, by comparing samples of active and non active galaxies, that the percentage of merging or gravitationally interacting systems is not higher in QSOs than in other types of active or non-active galaxies. However, this study, based on low resolution images ($1.7$arcsec/pxl), does not take into account faint tails or compact features in the host typical of recent merging activity. \citet{Li}, with a sample of $10^5$ low-redshift galaxies, find a strong correlation between the presence of a close companion and the star formation, but not between close companions and AGN activity. The puzzle is thus far from being solved.\\
In this framework, it is important (1) to find peculiar cases in specific stages of evolution, such as the famous HE0450-2958 \citep{Mag2}, which would allow to better assess the nature and creation mechanisms of the black hole-bulge coevolution \citep{Elbaz,Jahnke2009} and (2) to extract, in well resolved samples, correlations between observables as well as the proportion of interacting systems. 
In order to bring new insights on those issues, we study a sample of $69$ QSOs extracted from a complete sample of $104$ QSOs drawn from different catalogues. 
Section 2 explains the construction of the sample and the observations. Section 3 contains a description of how the contributions of the host and QSO are separated. The parameters derived thanks to this separation (QSO and host magnitudes, center of luminosity, asymmetry coefficient), the correlations found between them, and subsequent discussion are given in Sections 4 and 5. Some conclusions are drawn in Section 6. Finally, a few particularly interesting cases are analysed with more scrutiny in the Appendix. 

\section{Sample and observations}

\subsection{Sample}
We selected from the main catalogues available \citep{Veron,sdss,hewitt,eso,pg} all the brightest QSOs ($R\leq 18$) with $\delta \leq 10^{o}$ and $0.142\leq z \leq 0.198$. This redshift range was chosen to enable the observation of low redshift quasars through two specific filters, one including emission lines, the other only the QSO and stellar continua.  This provided a sample of $104$ QSOs, $9$ of which had already been observed at high resolution (see Table \ref{arch_sampl} for references), and $60$ of which were observed in the context of the present study.  Those $69$ QSOs form the sample analysed here. 
%The remaining $35$ have been observed with NTT/EFOSC2 (P082.B-0281 in october 2008 and february 2009, after SUSI2 decommissioning). However, the substancial and hardly predictable PSF variations across the field don't allow to achieve a QSO-host separation as reliable as for SUSI2. Consequently, they are not included in this paper. 
Even if only $66\%$ complete, we expect our sample to be statistically relevant and large enough to infer robust proportion measurments. Indeed, the $35$ remaining QSOs could not be observed only because of bad weather conditions during parts of the two runs, which is a selection independent of the QSOs characteristics, and thus devoid of any bias.  
\begin{table*}
 \centering
%  \caption{List of the QSOs already observed}
  \begin{tabular}{llrcrcrrrc}
  \hline
   Name     &   RA   & Dec & z  &  Filter & Ref. &   $M_V(QSO)$   & $M_V(host)$ & Morph. & Interac? \\
  \hline 
  \hline
   PHL909  &  00 54 32 & 14 29 58 & 0.171  & F606W & (1)&  -23.64 & -21.74 & Ell.&  N  \\
   0205+024 & 02 05 14.53 & 02 28 42.7 & 0.155  & F606W & (1)& -23.74 & -19.84 & Spir.& N\\
   PKS 0736+017 & 07 36 42.49 & 01 44 00.1 & 0.191  & F675W & (2)& -23.97 & -22.78 & Spir & Y\\ 
   PKS 1020-103 & 10 20 04.2 & -10 22 33.6 & 0.197  & F675W & (3)& ? & -21.29 & Ell. & N\\
   MC 1635+119 & 16 35 25.88 & 11 55 46.4 & 0.146  & F606W & (4)&-22.74 & ? & ? & Y\\
   PG 2349-014 & 13 49 22.3 &  -01 25 54 & 0.174  & F675W & (3)& ? &  -22.93 & ? & Y\\
   3C273 &  12 29 06.7 & 02 03 08 & 0.158 &  F606W & (1)&  -26.34 & -22.84 & Ell. & N\\
   HE1405-1545 & 14 08 24.5 & -15 59 28 & 0.194  & B & (5)& -24.2 & -23.3 & Spir. & Y\\
   HE1434-1600 & 14 36 49.6 & -16 13 41 & 0.144  & F606W & (6)& -23.82 & -22.87 & Ell. & Y\\ 
\hline
\end{tabular}
\caption{List of the QSOs already observed and main host properties. Reference: (1): \citet{Bahcall97}, (2): \citet{Dunlop}, (3): \citet{Kim}, (4): \citet{Canalizo2007}; (5): \citet{jahnke}; (6): \citet{LetaweY}. Objects from (1) to (4) were observed with HST/ACS/WFC, (5) with NTT/EFOSC2 and (6) with HST/ACS/HRC.  A question mark in $M_V(QSO)$ is when both absolute and apparent V-magnitudes are unknown.}
\label{arch_sampl}
\end{table*}

\subsection{Observations}
\begin{figure} 
\centering
\includegraphics[height=7cm,width=8cm]{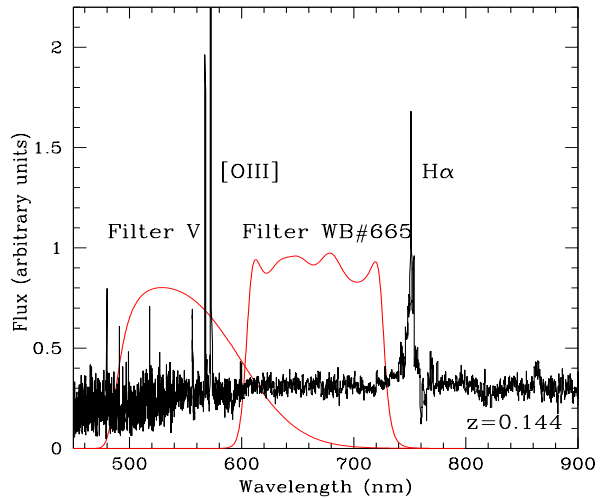}
\caption{VLT FORS1 spectrum of the host galaxy of the QSO HE$1434-1600$ after the separation from the QSO, at a redshift of $z=0.144$ \citep{Let3}, together with the response curves of the two filters used for probing the gaseous and stellar content.}
\label{filt}
\end{figure}

The observations of the sample was made with NTT/SUSI2, the Superb-Seeing Imager, a direct imaging camera optimised for periods of good seeing at the ESO/La Silla observatory, during the run A 077.B-0229 in August 2006 and run B 078.B-0081 in February 2007. We observed through two filters, V$\#812$ and WB$\#665$. The complete list of targets is given in Table \ref{sample}.
The observation strategy was motivated by two major points in the understanding of the QSO-host interactions. First of all, good resolution and sampling are necessary for detecting the host and revealing its morphology. SUSI2, in its $2*2$ binning mode, offers a sampling of $0.161$ arcsec/pxl, which translates in our redshift range into $\sim 0.45$ kpc/pxl. Exposure times have been estimated with the NTT ETC in order to reach a S/N high enough to detect galaxies departing by as much as $3\sigma$ from the mean magnitude relation between QSO and host. Each QSO observation was divided into $3$ or $4$ exposures in each filter in order to avoid saturation and to efficiently remove bad CCD pixels and cosmic ray hits. This set up, with a typical seeing of $0.6$, gives us observations deep enough to infer the major galactic morphology.  Non photometric observing conditions were reported for 8 objects.\\
Secondly, our previous study \citep{LetaweY} shows that the distribution of ionized gas in the host does not necesserally match the structure of the stars distribution . For instance, the host of HE0354-5500, analysed in \citet{LetaweY} contains gas ionized by the QSO even in remote regions devoid of stars. This motivates to study the gas and stellar content separately for each host. That is why all the $60$ SUSI2 QSOs have been observed through two filters: V$\#812$ which, in our redshift range, contains the two forbidden \ot $\lambda 4959$ and \ot$\lambda 5007$ lines along with \hb  $\lambda 4861$. Consequently, this filter allows to map the morphology of the ionized gas in the host, as we demonstrate in Section \ref{coul}. The filter WB$\#665$ contains a region devoid of emission lines, in which the stellar continuum should thus be the only contributor, allowing to map the morphology of the stellar content of the host. The presence of the stellar continuum in the V$\#812$ and its influence on the observed intensities are discussed in Section \ref{coul}. The filters response curves are plotted along with a typical QSO host spectrum in Fig. \ref{filt}. All observations were flatfielded and cleaned from bad pixels using PYRAF tools. 
Throughout the paper, we adopt the following cosmological parameters: $H_0=71\kms$Mpc$^{-1}$, $\Omega_m=0.27$ and $\Omega_\lambda = 0.73.$

\begin{table*}
 \centering
 \begin{minipage}{140mm}
  \begin{tabular}{lrrrrrrrr}
  \hline
   Name    &   RA   & Dec & z & Run &  V (s) & WB (s)& Morphology & Interaction? \\
  \hline 
  \hline
   005709+144610.1 & 00 57 09.9 & 14 46 10  &  0.172  & A & 1500 & 1800 & Ell. & Y\\
   011110-101631.8 & 01 11 10.0 & -10 16 32 & 0.179 & A & 750  & 930 & Spir. & Y\\
   011845+133327.1 & 01 18 45.5 & 13 33 27 & 0.189 & A & 2100 & 2625& Ell. & N\\
   015530-085704.0 & 01 55 30.0 & -08 57 04 & 0.165 & A & 640  & 800 & Und. & Y\\
   021218-073719.8 & 02 12 18.3 & -07 37 20 & 0.174 & A & 1350 & 1650 & Spir. & N\\
   021360+004226.7 & 02 13 59.8 & 00 42 27 & 0.182 & A & 1350 & 1800 & Und. & Y\\
   025007+002525.3 & 02 50 07.0 & 00 25 25 & 0.198 & A & 1200 & 1600 & Und. & Y \\
   032214+005513.4 & 03 22 13.9 & 00 55 13 & 0.185 & A & 640 & 800 & Ell. & Y\\
   101044+004331.3 & 10 10 44.5 & 00 43 31 & 0.178 & B & 576 & 720 & Spir. & N\\
   113706+013947.9 & 11 37 06.8 & 01 39 48 & 0.193 & B & 1536 & 1720 & Spir. & N\\
   122534-024757.2 & 12 25 34.8 & -02 47 57 & 0.195 & B & 1260 & 1630 & Ell. & N\\
   161532-002730.3 & 16 15 32.3 & 00 27 30 & 0.146 & B & 1260 & 1310 & Ell. & N\\
   205032-070131.2 & 20 50 32.3 & -07 01 31 & 0.168 & A & 2400 & 2800 & Spir. & N\\
   231712-003603.6 & 23 17 11.8 & -00 36 04 & 0.186 & A & 1940 & 2320 & Und. & Y\\
   232260-005359.3 & 23 23 00.0 & -00 53 59 & 0.150 & A & 1260 & 1440 & Spir. & N\\
   235156-010913.3 & 23 51 56.1 & -01 09 13 & 0.174 & A & 400 & 500 & Ell. & N\\
   Q 0022-2044 & 00 25 08.4 &-20 27 35 & 0.170 & A & 240 & 300 & Ell. & N\\
   CT 289 & 01 00 39.5 & -25 38 28 & 0.158 & A & 600 & 800 & Ell. & Y\\
   MS 01325-4151 & 01 34 42.7 & -41 36 13 & 0.172 & A & 600 & 750& Und. & Y\\
   MS 10302-2757 & 10 32 36.1 & -28 13 26 & 0.148 & B & 600 & 750& Ell. & N \\
   PKS 1241-399 & 12 44 29.4 & -40 12 46 & 0.191 & B & 1816 & 2160& Und. & Y\\
   MS 13591+0430 & 14 01 36.7 & 04 16 25 & 0.163 & B & 1272& 1590 & Ell. & Y\\
   Q 1421-0013 & 14 24 03.8 & -00 26 58 &0.151 & B & 600 & 750 & Spir. & N\\
   PDS 456 & 17 28 19.9 & -14 15 56 & 0.184 & A & 60 & 60 & Und. & Y\\
   MS 20078-3622 & 20 11 08.8 & -36 13 10 & 0.177 & A & 420 & 520& Ell. & Y \\
   Q 2240-2411 & 22 43 40.9 & -23 55 16 & 0.184 & A & 1050 & 1200& Spir. & N\\
   Q 2252-2434 & 22 55 25.0 & -24 18 30 & 0.147 & A & 450 & 240& Und. & Y\\
   6QZ J232927-2938 & 23 29 27.4 & -29 38 47 & 0.193 & A & 560 & 720& Spir. & N\\
   HE 0027-3118 & 00 29 37.3 & -31 02 10 & 0.145 & A & 900 & 1125& Ell. & Y\\
   HE 0108-5422 & 01 10 38.4 & -54 06 40 & 0.186 & A & 750 & 900 & Ell. & N\\
   HE 0146-3755 & 01 48 21.1 & -37 40 20 & 0.147 & A & 600 & 750 & Spir. & N\\
   HE 0226-5209 & 02 27 59.6 & -51 56 32 & 0.145& A & 1200 & 1530 & Und. & Y\\
   HE 0227-4123 & 02 29 13.3 & -41 10 10 & 0.143 & A & 1200 & 0 & Ell. & Y\\
   HE 0250-4400 & 02 52 23.0 & -43 47 55 & 0.168 & A & 540 & 600 & Ell. & Y\\
   HE 0441-2826 & 04 43 20.7 & -28 20 52 & 0.155 & A & 150 & 180 & Ell. & Y\\
   HE 1101-0959 & 11 04 16.7 & -10 16 08 & 0.186 & B & 1060 & 1260 & Ell. & N\\
   HE 1202-0501 & 12 04 53.0 & -05 18 13 & 0.169 & B & 1332 & 1668 & Und. & Y\\
   HE 1211-1905 & 12 14 03.4 & -19 21 43 & 0.148 & B & 1590 & 1935 & Spir. & Y\\
   HE 1236-2001 & 12 39 01.7 & -20 17 30 & 0.196 & B & 1248 & 1500 & Ell. & Y\\
   HE 1255-0437 & 12 58 31.0 & -04 53 49 & 0.172 & B & 1104 & 1380 & Ell. & Y\\
   HE 1256-2139 & 12 59 02.4 & -21 55 38 & 0.146 & B & 1200 & 1500 & Ell. & Y\\
   HE 1300-0657 & 13 02 46.7 & -07 13 55 & 0.181 & B & 1470 & 1830 & Spir. & Y\\
   HE 2345-3939 & 23 48 12.1 & -39 23 07 & 0.196 & A & 1090 & 1125 & Spir. & N\\ 
   0056-363 & 00 56 15.8 & -36 22 17 & 0.162 & A & 800 & 1000 & Ell. & Y\\ 
   0132+077 & 01 32 31.7 & 07 43 47 & 0.147 & A & 1050 & 1320 & Spir. & Y\\
   0213-484 & 02 13 52.6 & -48 26 55 & 0.168 & A & 1080 & 1260 & Ell. & Y\\ 
   0357+107 & 03 57 27.1 & 10 46 48 & 0.182 & A & 544 & 680 & Ell. & Y \\
   1001+054 & 10 01 43.3 & 05 27 34.8 & 0.161 & B & 750 & 900 & Ell. & Y\\
   1012+008 & 10 12 20.8 & 00 48 33 & 0.185 & B & 684 & 855 & Und. & Y\\
   1023-014 & 10 23 03.9 & -01 24 45.4 & 0.150 & B & 2060 & 2350 & Und. & Y\\	 
   1047+067 & 10 47 00.8 & 06 45 15.0 & 0.148 & B & 1176 & 1470 & Ell. & N\\ 	 
   1047-281 & 10 47 55.3 & -28 07 45.0 & 0.190 & B & 576 & 720 & Ell. & N\\
   1151+117 & 11 51 15.7 & 11 45 10.0 & 0.176 & B &  366 & 456 & Ell. & Y\\
   1226+136 & 12 26 54.6 & 13 36 54.0 & 0.150 & B & 1700 & 2250 & Ell. & N\\	 
   1241+095 & 12 41 10.1 & 09 33 31.3 & 0.190 & B & 2064 & 2580 & Spir. & N\\	
   1307+085 & 13 07 16.2 & 08 35 47 & 0.155 & B & 684 & 884 & Ell. & Y\\
   1325-012 & 13 25 59.8 & -01 13 47.2 & 0.150 & B & 1032 & 1290 & Und. & Y\\
   1850-782 & 18 50 08.0 & -78 15 00.0 & 0.162 & A & 180 & 208 & Ell. & N\\
   2140-457 & 21 40 10.0 & -45 42 29 & 0.171 & A & 840 & 990 & Und. & Y\\
 \hline
\end{tabular}
\end{minipage}
\caption{The NTT/SUSI2 sample observationnal characteristics, along with host morphologies (spiral, elliptical, or undefined) from visual inspection. The presence of signs of interaction is also indicated in the last column.}
\label{sample}
\end{table*}
\section{Image analysis}

\subsection{NTT/SUSI2 data processing}

\subsubsection{Deconvolution method}

The whole analysis of host galaxy and QSO properties relies on a good separation of those two components. We use a deconvolution method based on the MCS algorithm \citep{Mag}, which is known to be particularly well suited to separate point sources from a diffuse background \citep{Letawe,Let3,LetaweY,Mag2}. Its principle is to produce images with an improved resolution without trying to reach an infinite one.  Other techniques, which deconvolve an image with its true point spread function (PSF) and attempt to reach arbitrarily high resolutions, tend to result in artefacts, deblending problems, flux errors, and astrometry errors in the deconvolved image.  In the MCS technique, the user chooses some finite value for the resolution of the deconvolved output image by defining an output PSF smaller than the original image's PSF.  The user then chooses an appropriate pixel scale for the deconvolved output image in order to sample the chosen PSF at least at the Nyquist level.  By fully sampling the chosen PSF, the MCS algorithm eliminates artefacts in the deconvolved images, improves deblending of point sources, and reduces flux and position errors.  For our purpose, the deconvolved image has a pixel size times smaller than in the original image and a PSF of Gaussian shape, with a Full Width at Half Maximum (FWHM) of $2$ CCD pixels ($4$ in the deconvolved images). Another advantage of the method is that the point sources have a known shape in the deconvolved image and can be explicitely separated from the diffuse components, which allows to get an image of the host galaxy uncontaminated by the QSO. A particularity of this method in the present context of QSO host analysis is that it introduces no model or prior assumption for the shape of the host, as the galaxy is represented by a numerical diffuse component.

\subsubsection{PSF construction}
\begin{figure}
\centering
 \includegraphics[height=8.cm,width=6cm,angle=-90]{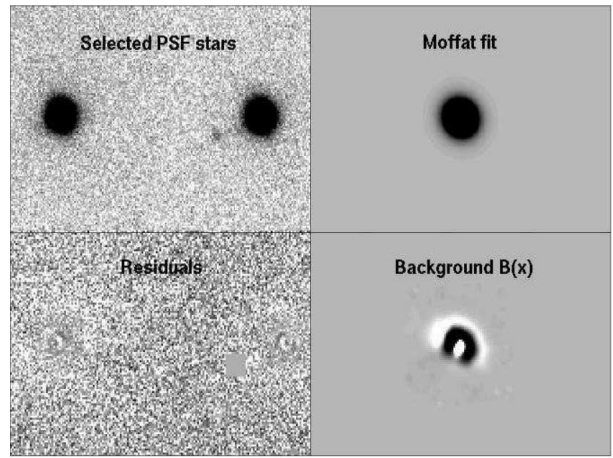}
\caption{Upper-Left: Different PSF stars are selected on the CCD. Upper-Right: A Moffat profile is adjusted on all the stars simultaneously. Bottom-Right: Numerical background added to improve the PSF. Bottom-Left: Residuals showing only faint structures particular to each star. }
\label{psf}
\end{figure}

A crucial point in the deconvolution process is the PSF construction: the more accurate the PSF at the position of the QSO on the CCD, the better the deconvolution. We construct the PSF as follows: 

\begin{itemize}
 \item Some stars (between $1$ and $4$) are selected on the same frame as the QSO, as close to it as possible and with similar brightness. Indeed, a star as luminous as the QSO has the same S/N and possibly suffers from the same deviations from linearity. Moreover, a star far from the QSO is more likely affected by PSF variations across the field, especially if located near the border. 
\item The kernel $s(\vec{x})$ of the deconvolution is first approximated by a Moffat profile (better reproducing the wings of the PSF than a Gaussian profile). Because real PSFs have more complex structure than this analytic function, we add to it a numerical component whose sole purpose is to improve the fit with real stars. This kernel (Moffat profile + numerical background), after convolution with the PSF of the deconvolved image (i.e.\  the chosen Gaussian profile) is then simultaneously fitted on the selected PSF stars.  A detailed description of the way the kernel is constructed can be found in \citet{Mag3}.  
\end{itemize}
An example of the whole PSF construction process is given on Fig. \ref{psf}. Only very faint structures, different for each star, are left in the residuals (differences between the model and the data for each pixel, divided by the standard deviation in this pixel). It shows that the relevant available information (i.e.\  common to all selected stars) is taken into account in $s(\vec{x}).$ 

\subsubsection{Simultaneous deconvolution}

After the construction of the PSF for each QSO exposure, the simultaneous deconvolution of the different images of a given QSO observed through a given filter can be carried out. The QSO image is decomposed into its point source component and a numerical diffuse background identical for all exposures, thus containing the full available information on the host. 

\begin{figure}
\centering
 \includegraphics[width=8cm]{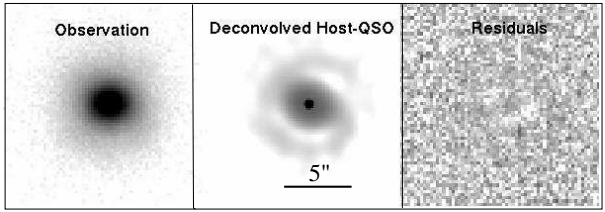}
\caption{Left: One of the $3$ exposures of the QSO 6QZJ2329-2938. Middle: The background and the point source at its fixed 2 pixels resolution.  Right: Residuals of the whole process, i.e.\  the reconvolved model minus the observation.}
\label{decsim}
\end{figure}

The parameters to fit are thus: the intensity and center of each point source, the common numerical background, which can be renormalised between different exposures if, for example, exposure times differ, and a global shift between images.
An example of the whole process of simultaneous deconvolution is given Fig. \ref{decsim}. It clearly shows that simultaneous deconvolution allows to: (1) seperate efficiently the QSO and its host, as the residuals are very good; (2) reveal the morphology of the host.\\ 
Among the whole observed sample, only $2$ QSOs, namely MS13591+0430 and 231711-003604, could not be efficiently deconvolved, because they fall very close to the line of sight of a very bright star. They were thus removed from the photometric analysis, even if visual inspection has given indications on their host morphologies.\\
We can now use the information provided by the deconvolution to compute some physical parameters such as the proportion of spiral and elliptical hosts, interacting systems, the host and QSO magnitudes in both filters, the center of luminosity of the host, its asymmetry coefficient, and try to find some correlations between those parameters. 

\subsection{Archive data}

As mentioned earlier, $9$ QSOs were already observed at high resolution, which allow to determine their host morphology and magnitudes. If magnitudes are given in the B or R band in the original paper, host V magnitudes are deduced from galaxy colours tables found in \citet{Fuku}, and QSO V magnitudes are deduced from apparent V magnitudes taken from the NED database, if available. They are upper limits as they include the host luminosity. A summary of these additional data is given in Table \ref{arch_sampl}. 

\section{Analysis}
Deconvolution of observations obtained in both filters provides separated host and QSO images, thus allowing the computation of different parameters, which we review one by one.

\subsection{Magnitudes}

\subsubsection{Apparent magnitudes}

The QSO flux is obtained as an output of the MCS algorithm. In order to avoid overestimation of the host flux, we flagged out the neighbouring objects lying in the field of view of the deconvolved image, by visual inspection. The  host flux is then computed by summing the intensities of the pixels of the deconvolved image. The total flux of each component can be converted in apparent magnitudes. This requires the knowledge of the apparent magnitude of a standard star. Observations of the TPHE B standard K7-star from the Landolt standard stars catalogue \citep{lando} allow to infer the zero-point in the V-band, but the apparent magnitude is not given for that star in the less common WB\#665 filter. However, it spans a similar wavelength range as the usual R-band. Thus, we deduce the star WB\#665 apparent magnitude by taking the fluxes of Vega and of a template K-star spectrum through the R-filter and through the WB\#665 filter:
\begin{equation}
m_{R}-m_{WB}=-2.5\log{\dfrac{\frac{Fl_{\rm{Kstar}}*R}{Fl_{\rm{Vega}}*R}}{\frac{Fl_{\rm{Kstar}}*WB}{Fl_{\rm{Vega}}*WB}}}=-0.03.
\end{equation}
Apparent magnitudes in both filters are listed in Table \ref{mag}. 
$8$ QSOs suffer from significant flux variations between each exposure, indicating non-photometric observing conditions. This led us to consider the apparent magnitude taken from the brightest exposure in each filter as an upper limit for the magnitude (lower brightness limit). 

\subsubsection{Absolute magnitudes}

The cosmic recession velocity of the objects we observe implies a shift in wavelength between the extragalactic emission and the observation we make of it. The appropriate correction, called K-correction, translate the observed magnitude measured through a filter to the intrinsic emitted magnitude, given the redshift of the object considered. As the host and QSO spectra are different, the conversion of apparent to absolute magnitudes uses different K-corrections for each component. Host galaxies are K-corrected taking into account their morphologies and redshift according to \citet{pence}, Table 14, which, although rather old, proved to be very reliable \citep{Kinney}. QSOs K-corrections are taken from \citet{crist}. They are only available in the V-band, thus WB\#665 absolute magnitudes could not be computed. 
V absolute magnitudes are given in Table \ref{mag}.

\begin{table*}
 \centering
% \begin{minipage}{140mm}
%  \caption{The NTT/SUSI2 sample magnitudes and asymmetry coefficient.}
  \begin{tabular}{lcccccccccccr}
  \hline
       &  \multicolumn{7}{c}{Magnitudes} & \multicolumn{5}{c}{Asymmetry} \\
  \hline
   Name&\multicolumn{3}{c}{QSO}&&\multicolumn{3}{c}{Host}&&\multicolumn{2}{c}{$\;a_{QSO}$}&\multicolumn{2}{c}{$a_{Gal}$} \\
           &  $m_V$  & $M_V$ & $m_{WB}$&& $m_V$  & $M_V$ & $m_{WB}$ && V & WB & V & WB\\
  \hline\hline
   005709+144610.1 & 15.31 & -24.59 & 15.26 && 16.95 & -22.49 & 16.34 && 0.84 & 0.92 & 0.26 & 0.8\\
   011110-101631.8 & 15.94 & -24.02 & 15.81 && 16.87 & -22.84 & 16.29 && 0.7 & 0.64 & 0.27 & 0.21\\
   011845+133327.1 & 17.49 & -22.68 & 17.24 && 16.97 & -22.97 & 16.28 && 0.13 & 0.12 & 0.14 & 0.1\\
   015530-085704.0 & 16.26 & -23.46 & 16.16 && 17.02 & -22.6  & 16.55 && 0.23 & 0.91 & 0.38 & 0.53\\
   021218-073719.8 & 17.16 & -22.70 & 17.17 && 17.55 & -21.84 & 17.19 && 0.05 & 0.34 & 0.12 & 0.12\\
   021360+004226.7 & 14.55 & - & - && 19.54 & - & - &&1.34 & 1.54 & 0.34 & 0.19\\
   025007+002525.3 & 16.93 & -23.44 & 16.87 && 18.69 & -21.57 & 18.01 && 0.91 & 0.91 & 0.23 & 0.13\\
   032214+005513.4 & 15.6 &  -24.71 & 15.52 && 17.87 & -21.92 & 17.27 && 0.64 & 0.77 & 0.15 & 0.35\\
   101044+004331.3 & 15.6 & -24.33  & 16. && 17.52 & -22.17 & 16.7 && 0.06 & 0.16 & 0.16 & 0.07\\
   113706+013947.9 & 16.22 & -23.89 & 16.23 && 18.34 & -21.52 & 17.64 && 0.15 & 0.65 & 0.26 & 0.8 \\
   122534-024757.2 & 16.93 & - & 16.85 && 18.39 & - & 17.86 && 0.41 & 0.09 & 0.16 & 0.22 \\
   161532-002730.3 & 16.7 & -23.00 & 16.47 && 16.71 & -22.57 & 16.43 && 0.1 & 0.53 & 0.08 & 0.08\\
   205032-070131.2 & 16.95 & -  &16.96 && 16.75 & - & 16.1&& 0.24 & 0.39& 0.15& 0.28\\
 %  231712-003603.6 & & & & & &\\
   232260-005359.3 & 16.73 & - & - && 16.56 & - & - && 0.74 & 0.71 & 0.45 & 0.17\\
   235156-010913.3 & 14.55 & - & - && 19.54 & - & - & &1.38 & 0.82 & 1.25 & 0.34\\
   Q 0022-2044 & 15.59 & -24.19 & 15.18 && 18.59 & -20.73 & 20.52 && 0.73& 1.45& 0.23 & 0.33\\
   CT 289 & 16.41 & -23.21 & 16.38 && 18.71 & -20.47 & 17.97 && 0.24 & 0.24 & 0.13 & 0.22 \\
   MS 01325-4151 & 17.54& -22.25& 17.21 && 17.53 & -22.16 & 16.94 && 0.30& 0.43&0.38 & 0.22\\
   MS 10302-2757 & 15.35 & -24.24 & 15.40 && 16.79 & -22.38 & 16.36&& 0.47 & 0.68 & 0.23 & 0.05 \\
   PKS 1241-399 & 17.96 & -22.43 & 17.74 && 17.67 & -22.60 & 16.88 && 1.02 & 1.1 & 0.65 & 0.64\\
%   MS 13591+0430 & & & & & &\\
   Q 1421-0013 & 15.65 & -23.94 & 15.64 && 17.47 & -21.91 & 17.64 && 0.54 & 0.29 & 0.13 & 0.23\\
   PDS 456 & 13.42 & -28.20 & 13.10 && 16.91 & -24.60 & 16.30 && 1.66 & 2.3 & 1.4 & 0.36\\
   MS 20078-3622 & 16.27 & -23.81 & 16.30 && 17.73 & -21.85 & 17.48 && 0.42 & 0.68 & 0.73 & 0.71\\
   Q 2240-2411 & 17.49 & -22.50 & 17.39 && 16.73 & -23.02 & 16.22 && 0.87 & 0.84& 0.18 & 0.28\\
   Q 2252-2434 & 15.44 & -24.01 & 15.55 && 16.89 & -22.47 & 16.15 && 0.97 & 0.87 & 0.33 & 0.14\\
   6QZ J232927-2938 & 15.85& -24.25 & 15.82 && 17.83 & -22.00 & 16.98 & &0.43 & 0.21 & 0.14 & 0.05\\
   HE 0027-3118 & 16.10 & -23.29 & 16.19 && 17.53 & -21.44 & 16.88 && 0.52 & 0.42 & 0.12 & 0.26\\
   HE 0108-5422 & 16.70 & -23.31 & 16.62 & &19.01 & -20.47 & 17.65 && 0.43 & 0.56 & 0.29 & 0.04\\
   HE 0146-3755 & 15.98 & -23.43 & 16.03 && 16.38 & -22.83 & 15.89 && 0.72 & 0.47 & 0.21 & 0.35\\
   HE 0226-5209 & 16.64& - & 17.26 && 16.84 & - & 16.74 &&1.12 & 1.21 & 0.66 &0.58\\
   HE 0227-4123 & 18.84 & - & - &&19.59 & - & - & &0.24 & - & 0.37 & - \\
   HE 0250-4400 & 15.57 & -24.16 & 15.56 && 17.35 & -21.92 & 17.06 && 0.7 & 0.34 & 0.21 & 0.07 \\
   HE 0441-2826 & 14.45 & -25.16 & 14.32 && 16.04 & -23.13 & 16.29 && 0.73 & 1.09 & 0.08 & 0.11\\
   HE 1101-0959 & 17.08 & -23.03 & 17.04 & &17.69 & -21.89 & 17.26 & &0.39 & 0.47 & 0.08 & 0.08\\
   HE 1202-0501 & 16.15 & -23.67 & 16.06 && 16.5 & -23.21 & 16.28 && 1.62 & 1.73 & 1.23& 1.22\\
   HE 1211-1905 & 16.45& -23.10 & 16.55 && 16.64 &  -22.70& 16.13 && 1.14 & 1.4 & 0.41 & 0.60\\
   HE 1236-2001 & 16.04 & -24.22 & 15.85 && 17.21 & -22.50 & 16.75 && 0.51 & 0.46 & 0.21 & 0.16\\
   HE 1255-0437 & 16.34 & -23.49 & 16.35 && 16.86 & -22.50 & 16.25 && 0.15 & 0.29 & 0.32 & 0.45\\
   HE 1256-2139 & 15.79 & -23.93 & 15.75 & &18.81 & -20.49 & 18.75 && 1.9 & 1.68 & 0.43 & 0.61 \\
   HE 1300-0657 & 16.53 & -23.48 &  16.48 && 16.82 & -22.94 & 16.22 && 0.32 & 0.27 & 0.13 & 0.04\\
   HE 2345-3939 & 16.61&-23.50 &16.65 && 16.37&-23.47 &15.83 && 1.17 & 1.14 & 0.47 & 0.38\\ 
   0056-363 & 14.83 & -24.81 & 14.9 && 17.31 & -21.9 & 19.27 && 1.49 & 1.53 & 1.55 & 1.54\\ 
   0132+077 & 17.37 & -22.18 & 17.4 && 16.96 & -22.39 & 16.32 && 0.29 & 0.29 & 0.06 & 0.08\\
   0213-484 & 16.92 & -22.85 & 16.92 && 17.67 & -21.64 & 17.09 && 0.56 & 0.17& 0.14 & 0.19 \\ 
   0357+107 & 15.81& - & -24.88& &18.21 & - &-21.98 && 0.85 & 0.71 & 0.06 & -\\
   1001+054 & 15.57& -24.12&15.69 && 17.76 & -21.85 & 17.31 && 0.53 & 0.8&0.24 & 0.37\\
   1012+008 & 15.15& -24.89& 15.15 && 16.45& -23.47 & 15.76 && 2.08 & 1.95& 0.85 & 0.6\\
   1023-014 & 18.85& -20.73& 18.89  &&17.49 &-22 &17.38 && 2.04 & 1.95 & 0.51 & 0.18\\	 
   1047+067 & 16.82& -22.65 & 16.84 & &17.12 & -21.82 & 16.49 && 0.09 & 0.34 & 0.09 & 0.12\\ 	 
   1047-281 & 15.27 & -24.99 & 15.31 && 17.54 & -22.18 & 16.76 && 0.28 & 0.33 & 0.1 & 0.48\\
   1151+117 & 15.61& -24.28& 15.64&& 18.79& -21.00& 17.59&& 1.51 & 1.37 & 0.4 & 0.32 \\
   1226+136 & 17.43& -22.08 & 17.07 && 17.06 & -22.02 & 16.61 && 0.18 & 0.38 & 0.26 & 0.22\\	 
   1241+095 & 16.81 & -23.24 & 16.79 && 17.06 & -22.74 & 16.73 && 0.74 & 0.65 & 0.18 & 0.07\\
   1307+085 & 14.79 & -24.84 & 15.29 && 17.42 & -21.78 & 17.75 && 1.29 & 0.28 & 0.29 & 0.15\\
   1325-012 & 16.95 & - & 16.96 && 17.24 & - & 16.36 && 1.59 & 1.55 & 0.19 & 0.18 \\
   1850-782 & 14.99 & -25.20 & 14.94 && 17.00 & -22.75 & 16.35 && 0.42 & 0.55 & 0.13 & 0.1\\
   2140-457 & 15.68 & -24.12 & 15.64 && 17.92 & -21.77 & 17.47 && 0.45 & 0.39 & 0.12 & 0.12\\
 \hline
\end{tabular}
\caption{Apparent and absolute magnitudes are given separately for the QSO and the host in the V-band, while only apparent magnitudes could be computed in the WB\#665-filter. For non photometric observations, absolute V-magnitudes are not displayed and apparent magnitudes must be taken as a lower limit. Asymmetries computed with respect to the QSO center and to the center of luminosity of the host are given in both filters (see text for details).}
%\end{minipage}
\label{mag}
\end{table*}  

\subsubsection{Errors}
\label{mager}
Errors in the computation of magnitudes are dominated by two components: the photon noise inherent to the observation, and the accuracy of the QSO-host separation. The photon noise is estimated by taking the square root of the mean QSO+host flux of our sample. That leads to a mean error of $\simeq 0.05\%.$
%\begin{equation}
 %\sigma_{\gamma}=\sqrt{3.7*10^6}=1924 \Leftrightarrow Flux=3.7\pm 0.002*10^6
%\end{equation}

To estimate the error on QSO-host separation, we first selected 3 representative QSOs with different quality deconvolution residuals: 1241+095, HE0146-3755, and Q2252-2434. For each of these QSOs, we varied the value of the QSO intensity until clear signs of bad PSF subtraction appear (clear hole or peak at the QSO position).  Signs of bad PSF subtraction begin to appear when we vary the QSO intensity by 1.1 to 7.0 per cent, suggesting that this is the level of uncertainty in the QSO magnitudes. Propagating these errors to magnitude errors leads to $\sigma_M\simeq0.02-0.05$.

Another way to test the error bars on galactic and QSO magnitudes obtained through deconvolution is to add a simulated QSO on an observed galaxy in one of the observed fields and run the deconvolution process on this new image. The input and output fluxes may then be compared to estimate the error bars on magnitudes. After running this procedure for several objects, we observe a deviation of at most 20\% for the host flux and 2\% for the QSO flux. Not surprisingly, the uncertainty for the host is stronger when the QSO intensity is high relative to the host. On average, given the nuclear/host ratios in the sample, we can consider an uncertainty of 0.1 mag for the host magnitude and 0.01 for the QSO magnitude. The errors coming from the QSO/host separation thus clearly tend to dominate the photon noise.

\subsection{Morphology}

One of the first natural steps in the study of QSO-host samples is to seek if there is a link between the host morphological type and the QSO activity. More precisely, the current open questions are: Do gravitational interactions trigger QSO activity? If yes, what are the physical processes at work? Are there other ways of triggering activity? Which ones? Once the nucleus is turned on, how does it evolve with its host?...  One way to answer these questions is to find classification criteria for host galaxies that enable to compare them with quiescent galaxies. The bimodal distribution of galaxy colour-magnitude diagrams \citep{Baldry} allows such a comparison. Namely, \citet{Martin} find that the host galaxies of AGNs lie more often in between the red (corresponding to early-types) and blue (corresponding to late-types) sequence of such diagrams, suggesting that the AGN phenomenon is a transitionary step between those sequences. Another possibility is to simply classify hosts by their morphological type and find out the frequencies of interacting, elliptical and spiral systems, in which QSOs are found to lie, and compare them with previous results. \\
We chose to classify each host galaxy according to its morphology by visual inspection. Namely, the different selection properties are: (1) Spiral, elliptical or undefined morphology (exclusive), where the undefined morphology most often corresponds to cases in which the merging process is violent enough to totally disturb the host, preventing the determination of its morphology, and, (2) signs of gravitational interaction and presence of more than one point source (non-exclusive). The percentages, calculated on the total sample (SUSI2+archive sample), can be found in Table \ref{pourc}.
 
\begin{table}
 \centering
 \begin{minipage}{140mm}  
  \begin{tabular}{lrrrr}
  \hline
        &Regular&Interaction&Multiple nuclei&Total\\
  \hline 
  \hline
  Spiral& 17 & 8& 0 &  25\\
Elliptical& 23 & 28 & 4 & 51\\
Undef.& 0 & 24 &  4 &  24 \\   
 \hline
\hline
Total & 40 & 60 & 9 & \\
\hline
\end{tabular}
\end{minipage}
\caption{Occurency (in $\%$ of the whole sample) of morphology classes and interaction features.}
%\caption{Classification of the host morphology}
\label{pourc}
\end{table}

The proportion of ellipticals ($51\%$) and spirals ($25\%$) is essentially compatible with some previous results, such as \citet{Shade} or \citet{Letawe}. However, we find a lower proportion of elliptical galaxies than in \citet{Dunlop}, \citet{Hamilton} and \citet{Floyd}, but this may be due to a selection effect in their sample, as mentioned in \citet{Letawe}. 
Concerning the proportion of hosts showing signs of gravitational interaction, we just mention for the moment that the $60\%$ we find are much higher than the $5$ to $20\%$ found for quiescent galaxies in some previous studies \citep{Lefevre,Bell,Zheng}, although \citet{Shi} argue in favour of a higher merger rate.  One difficulty in estimating the proportion of merger remnants is that, as shown by simulations, the timescales during which they show clear signs of disturbance are often shorter than the total merger timescale \citep{lotz}. On the other hand, we find a proportion of interaction very similar to \citet{Schmitt} for both active and inactive galaxies. Namely, $\simeq55\%$ of all our elliptical hosts show signs of gravitational interactions, and this proportion drops to $\simeq32\%$ for spirals, while they find respectively $\simeq50\%$ and $\simeq25\%$. If we consider our \textit{undefined} class as strong mergers, our percentage of $24\%$ is very similar to \citet{Greene}, who find a proportion of one quarter for highly disturbed morphologies in obscured active galaxies. 

In conclusion, we can state that our estimates show that the morphologies of active galaxies are not different from the quiescent ones, and that the occurence of merger signs in QSO hosts is higher or similar to inactive galaxies, regarding previous discordant studies. If higher, we could assess the influence of mergers on activity. Thus detailed studies on the presence of merger signs in inactive galaxies at similar spatial resolution would be required before one can firmly conclude.
 
All in all, the major difficulty in comparing different studies lies in the definition and quantification of the degree of interaction in a system. It is often done by simple eye-check, which is always subjective. That leads us to define the following asymmetry coefficient.

\subsection{Asymmetry}
An idea that has been recurrent for the past 10 years is to quantify interactions and mergers via the asymmetry of the system. 
Several methods can be found in the literature. We briefly review the most relevant ones and their results. First of all, \citet{Abraham,Conselice1,Conselice2} and \citet{Shi} used an asymmetry index defined essentially as the subtraction of the image of a galaxy by the same image rotated by $180^{o}$. This simple method proved to be a good tracer of the degree of interaction in merging galaxies. With the help of two supplementary parameters, the concentration index and the clumpiness,  \citet{Conselice2} achieve to describe all usual morphological types in a totally quantitive way, solely from image analysis. Another method called Zurich Estimator of Structural Types (ZEST) has been proposed by \citet{Scarlata}. It uses the Sersic index $n$ returned from a galaxy fit and $5$ other basic nonparametric diagnostics (amongst which the asymmetry coefficient) to quantify the properties of galaxy structure, using Principal Component Analysis. They find strong evolutionary effects between $z=0$ and $z=0.7$ for faint galaxies ($M_B>-20.5$). \\
Another morphological parameter used as a signpost of a nonequilibrium global dynamical state is the so-called lopsidedness. It is defined as the radially averaged $m=1$ azimuthal Fourier amplitude measured between radii enclosing $50\%$ and $90\%$ of the galaxy light. Using this definition for a low-redshift sample ($z<0.06$) of $\sim25000$ galaxies, \citet{Reichard} find a trend for more powerful AGN to be hosted by more lopsided galaxies, and conclude that is due to the strong link of the age of the stellar population in the galaxy bulge to both the AGN luminosity and to the lopsidedness of the host.\\
These analyses concentrate on Type $2$ AGNs or quiescent galaxies, and do not take into account Type 1 QSOs, which considerably overweight the central regions of the host in the computation of the asymmetry index. In order to include QSOs in that kind of analysis, \citet{Gabor} first separe QSO and host components with the GALFIT tool, and then compute the asymmetry index in the same way as \citet{Conselice1}. They find that, at $0.3<z<1.0$, QSO host asymmetries are no more prevalent than in quiescent galaxies. 
Another way of computing asymmetries for QSO host galaxies is proposed by \citet{Kim}, using an all-GALFIT computation. After separing the QSO and the host, Fourier components are fit to the images showing significant nonaxisymmetric features. Their result is that, for low redshift QSOs ($z<0.35$), galaxy mergers and tidal interactions seem to play an important role in regulating and fueling the nuclear activity. The apparently opposite results in different redshift ranges from \citet{Gabor} and \citet{Kim} are suggestive of a cosmological evolution of the importance of interactions. It reinforces our belief that the study of asymmetries is a powerful tool to understand QSO-host interactions.\\
Here, in order to get rid of any assumption concerning the host morphology or the estimated center of the host, we propose an alternative way to define the asymmetry, inspired by the statistical third order moment, called skewness indice.
More precisely, if $I(\vec{x})$ is the intensity in the pixel $\vec{x}$, we define the center of luminosity of the host by
\begin {equation}
 \vec{x}_{c_L}=\dfrac{\sum_{\vec{x}} I(\vec{x})\vec{x}}{\sum_{\vec{x}} I(\vec{x})}.
\end {equation}
where the sums are over the total number $N$ of pixels in a square subimage defined by the user and containing the host galaxy.\\
 Then, we define an asymmetry coefficient with respect to a center $c$ located at some position $\vec{x}_c$  
\begin{equation}
 a_c=\dfrac{\sum_{\vec{x}} I(\vec{x})(\vec{x}-\vec{x}_c)^3}{\sum_{\vec{x}} I(\vec{x})}.
\end{equation}
This method has the advantage of being totally model independent, and takes into account the whole light from the host. It thus requires no a priori knowledge about the shape of a galaxy, which involves the risk to misguide any interpretation. \\
Another advantage of this definition is that $c$ can be chosen to be either the center of luminosity $c_L$, thus describing the asymmetry of the galaxy alone, or the QSO position $c_{{\rm QSO}}$, thus giving information about the asymmetry of the whole system (QSO+host). From now on, unless clearly stated otherwise, the term ``asymmetry coefficient'' will refer to the asymmetry with respect to $c_{{\rm QSO}}$. Indeed, a QSO located off-center in an otherwise seemingly symmetrical galaxy can be considered as a sign that something special is happening. For example, in the framework of binary black hole mergers, the resulting black hole is supposed to recoil in a direction opposite to the gravitational wave emission \citep{Redmount}. During this recoil, the accretion disc may stay bound to the black hole and consequently shine off-centered. Simulations \citep{Volonteri2008} predict that a population of off-nuclear AGNs may already be detectable at low and intermediate redshifts.\\ 

\begin{figure}
\centering
\includegraphics[height=4.8cm, width=8.2cm]{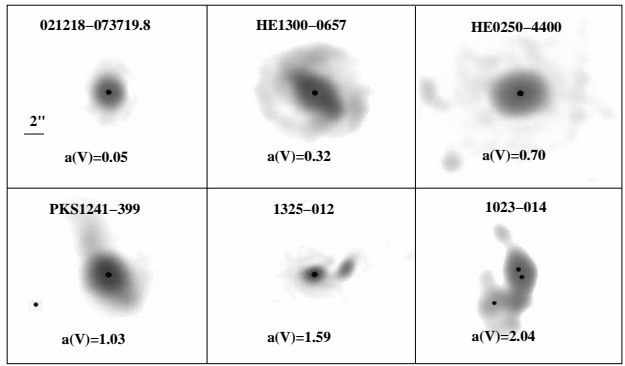}
\caption{Asymmetry coefficients in the V-band $a(V)$ are displayed for $6$ QSO hosts which are at different levels of interaction. A higher asymmetry value corresponds to a higher degree of interaction. The spatial scale is identical in all frames.} 
\label{asym}
\end{figure}

In order to minimize the importance of border effects linked to the choice of the square subimage where the asymetry coefficient is computed, we multiply the region of interest by a broad Gaussian function, centered on the center of luminosity.  A FWHM of $30$ pixels, matching the extent of the galaxy, reveals to be appropriate in the majority of cases as it is not too narrow, which would give more weight to asymmetries due to shifts between $c_L$ and $c_{{\rm QSO}}$ in comparison with galactic asymmetries, and not too broad to be useless because including unrelated features in the field. The FWHM was only changed when the galaxy was significantly smaller or larger (the extreme values for the sample being down to 15 and up to 45 pxl). If the width of the Gaussian is suited to the galactic extend, a difference of 2pxl FWHM does not induce significant changes in the asymmetry. 
 In Fig.~\ref{asym}, six deconvolved images of QSOs and their hosts are shown along with their asymmetry coefficient. This illustrates the relevance of using the asymmetry coefficient as a measure of the degree of interaction of a system. 

 The global sample can be divided into two subsambles, one with $a(V)\leq0.7$ (low asymmetry), and the other with $a(V)>0.7$ (high asymmetry). This value seems appropriate (see Fig. \ref{asym}) for a good separation between minor interaction events and galactic-scale mergers involving several galaxies. The low-asymmetry subsample contains $60\%$ of the sample.  Let us mention that asymmetry coefficients were not computed for the archive data because hosts were not separated from their nucleus. They are thus not included in forthcoming analysis including asymmetries.
\subsubsection*{Errors}

Errors on the asymmetry coefficient and $c_L$ have been estimated by varying the input parameters (size of the square subimage in which we compute the coefficient and Gaussian FWHM) and see how it influences the result. First, it shows that $c_L$ is very stable (a $10\%$ variation of the subimage size leads to a $<1\%$ variation of $c_L$). Conversely, the asymmetry coefficient is quite sensitive to the Gaussian FWHM (a $25\%$ variation in FWHM leads to a $\simeq 15\%$ variation in asymmetry coefficient). However, the restrictions for setting the FWHM value explained above do not allow such large variations, and thus we do not expect a change of more than $>10\%$ in the asymmetry coefficient, which is clearly sufficient for the present purpose. \\
Two other sources of error must be taken into account: the variation in sampling due to the redshift range spanned, and S/N variations, which might change the influence of the background noise on the asymmetry. \\
First of all, the exposure times were chosen to reach a comparable S/N for all the hosts, thus considerably lowering the risk of error due to S/N variations. 
Secondly, it might be expected that higher-$z$ hosts would show a lower asymmetry than their lower-$z$ counterparts, as the hosts substructures are less and less resolved with increasing redshift. In order to test the importance of this effect in our sample, we use the PYRAF tool ``magnify'' to decrease the resolution of the observations of the most asymmetric hosts ($a(V)\sim0.6$) among the nearest objects ($z<0.15$) from around 0.4 kpc/pxl to a 0.52 kpc/pxl scale, corresponding to the upper limit of the sample, $z=0.198$, and compute the asymmetry on this new image. The result is that the asymmetry coefficient drops at most by 0.02, revealing that the redshift range is small enough to avoid such biases in the estimates. We can thus neglect both of these effects.
The calculated asymmetries with respect to $c_{\rm{QSO}}$ and $c_L$ are given in Table \ref{mag} for both filters.

\section{Results}

Now that the parameters have been introduced, we aim to find correlations between them and try to specify some aspects of QSO-host interactions. 

\subsection{QSO-Host absolute magnitudes relation}

\begin{figure}
\centering
 \includegraphics[height=7.5cm,width=8cm]{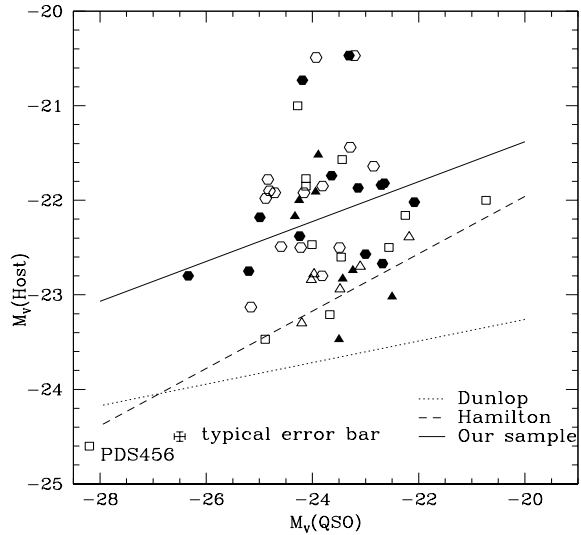}
\caption{$M_V(Host)$ is plotted versus $M_V(QSO)$. Solid line: the best fit relation of our sample. Dashed line: fit from \citet{Hamilton}. Dotted line: fit from \citet{Dunlop}. Filled symbols represent non-interacting galaxies, while open ones are for galaxies showing signs of interaction. Triangles are for spirals, hexagones for ellipticals, whereas squares stand for unclassified morphologies.}
\label{magabs}
\end{figure}

A common question in QSO hosts studies is to seek if there is a correlation between QSO and host absolute magnitudes. 
Theoretically, such a trend is expected as a consequence of the more fundamental and well constrained  $Mass(Spheroid)\textrm{-}Mass(Black\ Hole)$ relation \citep{Ferrarese,Marconi}. However, it might be strongly disturbed by the variety of accretion rates found in AGN, which prevents from linking the black hole masses and magnitudes unequivocally. Equally, spheroid masses are not expected to match the galaxies magnitudes, as the latter contain significant supplementary structures (tidal tails, spiral arms and bars, discs). 
Previous results \citep{Hamilton,Floyd,Shade} tend to indicate only a slight correlation, while others \citep{Bahcall97,Dunlop} find a relation compatible with no correlation at all. 
%An interesting discovery was made by \citet{Floyd} who find that, for a given host magnitude, the nucleus does not radiate stronger than its Eddington accretion limit, even if super-Eddington limit accreting QSOs have already been found, and are thought to correspond to an early stage of the QSO evolution \citep{Letawe,Hamilton}.\\
Our sample shows a weak correlation between host and QSO magnitudes (Fig. \ref{magabs}, where our correlation is compared with mean relations from other samples, and the morphology classification is indicated).  Namely, we find that, if $<M_V(QSO)>$ is the mean value of the whole sample,
\begin{eqnarray}
 M_{V}(Host)&=&(0.21\pm 0.10)(M_V(QSO)-<M_V(QSO)>) \nonumber\\
            &-&(22.20\pm 0.11),
\end{eqnarray}
with a Pearson's correlation coefficient $p=0.28$ (the correlation is calculated with $M_V(QSO)-\textrm{\textlangle}M_V(QSO)\textrm{\textrangle}$ instead of $M_V(QSO)$ so that the intersection with the $M_V(Host)$-axis corresponds to the average case where $M_V(QSO)=\textrm{\textlangle}M_V(QSO)\textrm{\textrangle}$). 

In Table \ref{corr}, the relations found for different subclasses are presented with associated correlation coefficients, all globally weak. 
% We see that the relation for non interacting spirals is perpendicular to the total sample one, which is thus mainly created by the ellipticals. It can be interpreted as an indication that the total $M_V(Host)-M_V(QSO)$ relation is only a remnant of the $Mass(Spheroid)-Mass(Black\ Hole)$ relation, which is strongly disturbed by the presence of supplementary structures in spiral galaxies.  However, it is hard to firmly conclude on that point as the deviations from the fit are quite large, and the correlations are often weak. In fact, we should not underestimate the fact that the QSO PDS456, which lies at the very high luminosity part of the graph, is likely to play a crucial role in the overall trend, but does not influence the trend found for the ellipticals and the spirals separately. 
Brighter ellipticals most probably harbour brighter QSOs, as theoretically expected, but this does not hold for spirals, resulting in a poor correlation between $M_V(Host)$ and $M_V(QSO)$ for the whole sample, in agreement with the aforementioned  previous studies.

\begin{table}
 \centering
 \begin{minipage}{140mm}
%  \caption{List of the QSOs already observed}
  \begin{tabular}{lrrr}
  \hline
   Subclass     &   $a$   & $b$ & $p$\\
  \hline 
  \hline
   All &    $0.21\pm0.10$ & $-22.20 \pm 0.11$ &  $0.28$ \\
   Spirals & $-0.28 \pm 0.39$ &  $-22.35 \pm0.26$ & $0.20$\\
   Spirals w/o inter. & $-0.5 \pm 0.61$ &  $-22.26 \pm 0.32$ & $0.26$\\
   Ellipticals & $0.23\pm 0.14$ & $-21.94\pm 0.13$ & $0.31$ \\
   Ellipticals w/o inter. & $0.17\pm 0.17$ & $-22.02\pm 0.20$ &  $0.29$\\
   Interactions & $0.26 \pm 0.12$ & $-22.22 \pm 0.14$ & $0.39$ \\
 \hline
\end{tabular}
\end{minipage}
\caption{Linear $M_V(Host)=a(M_V(QSO)-\textrm{\textlangle}M_V(QSO)\textrm{\textrangle})+b$ relation for different morphological subclasses. $p$ is the Pearson's correlation coefficient.}
\label{corr}
\end{table}

\begin{figure}
\centering
\includegraphics[height=8.5cm,width=8.cm]{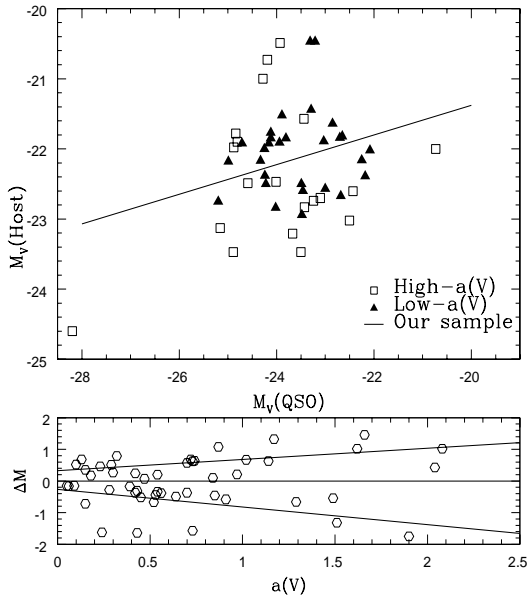}
\caption{Top. $M_V(Host)$ is plotted versus $M_V(QSO)$ with the QSOs labelled according to their asymmetry subsample. Bottom. Deviations $\Delta M$ from the best fit are plotted against the asymmetry coefficient, with a linear fit on the postive and negative values of $\Delta M$ separately.}
\label{magasym}
\end{figure}

As explained above, the asymmetry coefficient is a good indicator of the degree of disturbance due to interaction in a system. Thus, we can replace the traditional morphology classification used in Fig. \ref{magabs} by an asymmetry-based classification. Figure \ref{magasym} shows the $M_V(Host)-M_V(QSO)$ relation, with this new subsample separation.  

We find that the high-asymmetry subsample shows a dispersion twice larger around the fitted relation. This trend is even clearer when we plot the deviations from the fit as a function of the asymmetry (Fig. \ref{magasym}, bottom). It shows a correlation between both variables indicating that the higher asymmetry hosts deviate more strongly from the fit. \\
Moreover, the asymmetric hosts tend to lie on average below the fit in Fig. \ref{magasym}. It means that, for a given nucleus magnitude, highly asymmetric systems tend to have brighter host galaxies. It can simply be explained by the fact that important mergers, corresponding to high $a(V)$ values, contain two or more galaxies which boost the value of the computed magnitudes. More interesting are the systems which have a surprisingly low-luminosity host (or powerful nucleus) along with highly asymmetric features (open squares at the top of Fig. \ref{magasym}). Most probably, they are systems where interactions have most strongly enhanced the QSO activity and luminosity, as suggested for example in \citet{Letawe}.

\begin{figure}
\centering
 \includegraphics[width=7.cm]{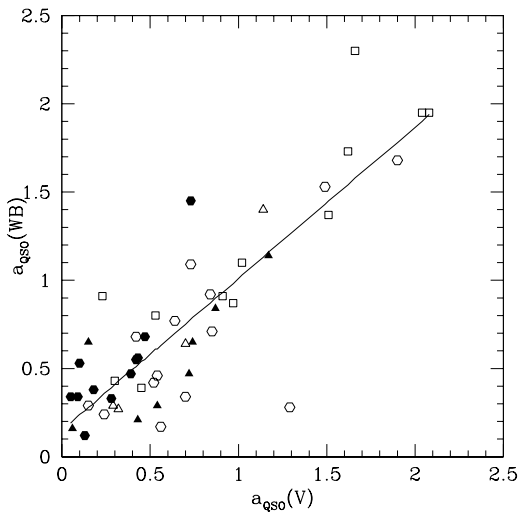}
\caption{Correlation between the asymmetry coefficients measured in both filters. Note that all the high asymmetry systems have also been classified as interacting systems. Filled symbols represent non-interacting galaxies, while open ones are for galaxies showing signs of interaction. Triangles are for spirals, hexagones represent ellipticals, whereas squares stand for unclassified morphologies.}
\label{asymVW}
\end{figure}

It is also interesting to check if asymmetric hosts in a filter are also asymmetric in the other one. Figure \ref{asymVW} shows that for the vast majority of the QSOs observed, asymmetries are, as expected, comparable in both filters. The best fit is $a(WB)=(0.86\pm0.08)a(V)+(0.15\pm 0.07)$, with a correlation coefficient $p=0.85$. However, some objects lie way out of the mean relation. They will be discussed individually in the Appendix.
Figure \ref{asymVW} is useful to describe the link between the two classification methods (morphology vs asymmetry). It is clear that the most disturbed cases, morphologically undefined and indicated by an open square, also have the highest asymmetry value. Spirals and ellipticals have comparable asymmetries, whatever the filter. Table \ref{asym_morph} summarizes the mean asymmetries in both filters as a function of morphology.

\begin{table}
 \centering
 \begin{minipage}{140mm}
%  \caption{List of the QSOs already observed}
  \begin{tabular}{lrrrr}
  \hline
   Subclass     &   $\textrm{\textlangle}a(V)\textrm{\textrangle}$ &$\sigma(V)$&  $\textrm{\textlangle}a(WB)\textrm{\textrangle}$ &$\sigma(WB)$ \\
\hline
\hline
 Spirals	&	0.59 &	0.36 & 0.58 & 0.39\\
 Ellipticals	&	0.57 &	0.45 & 0.61 & 0.42\\
 Interactions	&	1.11 &	0.66 & 1.23 & 0.63\\
 \hline
\end{tabular}
\end{minipage}
\caption{Average asymmetries are given in the two filters for spirals, ellipticals, and hosts with signs of interaction.}
\label{asym_morph}
\end{table}

%%%%%%%%%%%%%%%%%%%%%
\begin{figure*}
 \centering
\includegraphics[height=5.5cm]{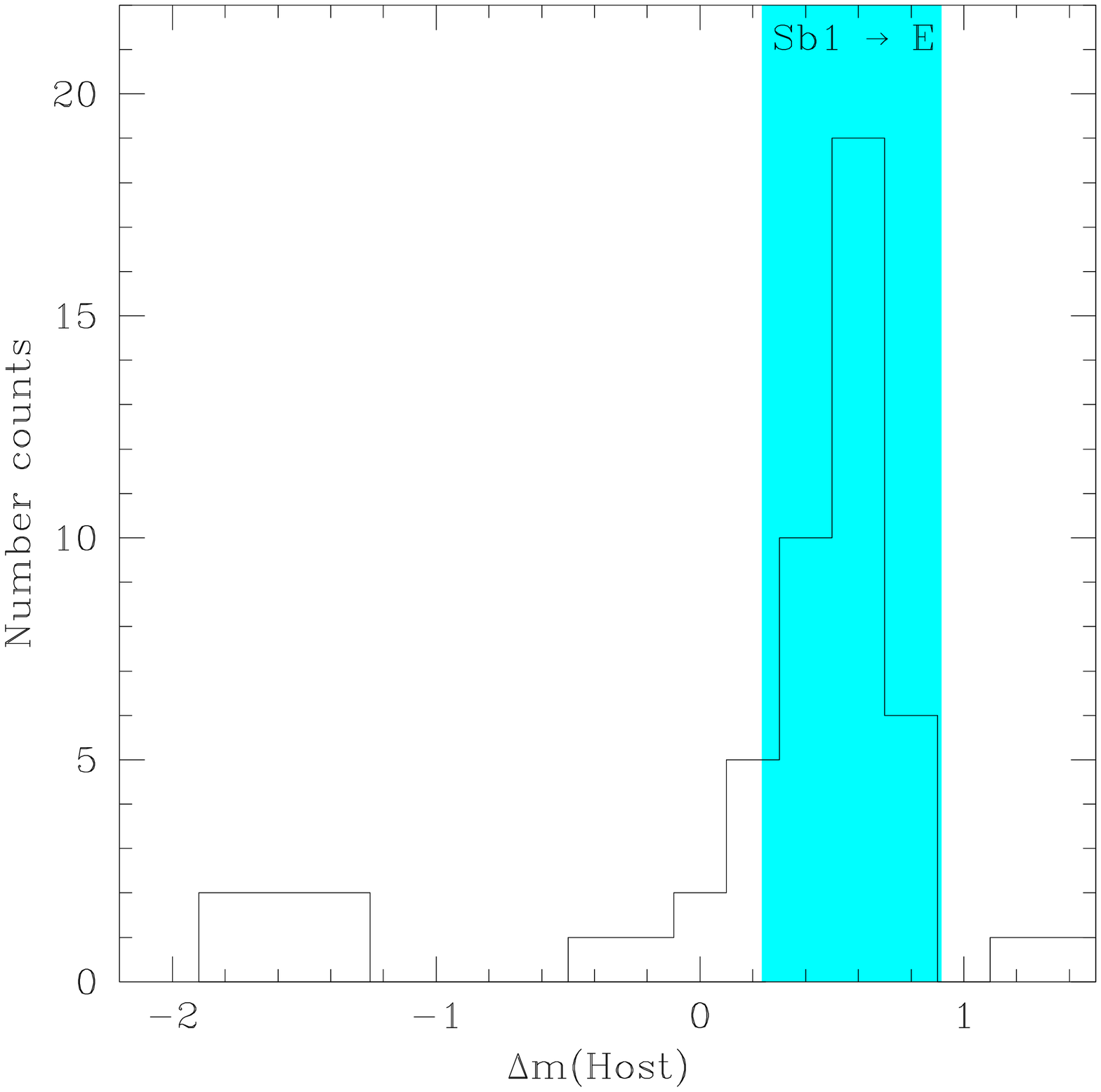}
\includegraphics[height=5.5cm]{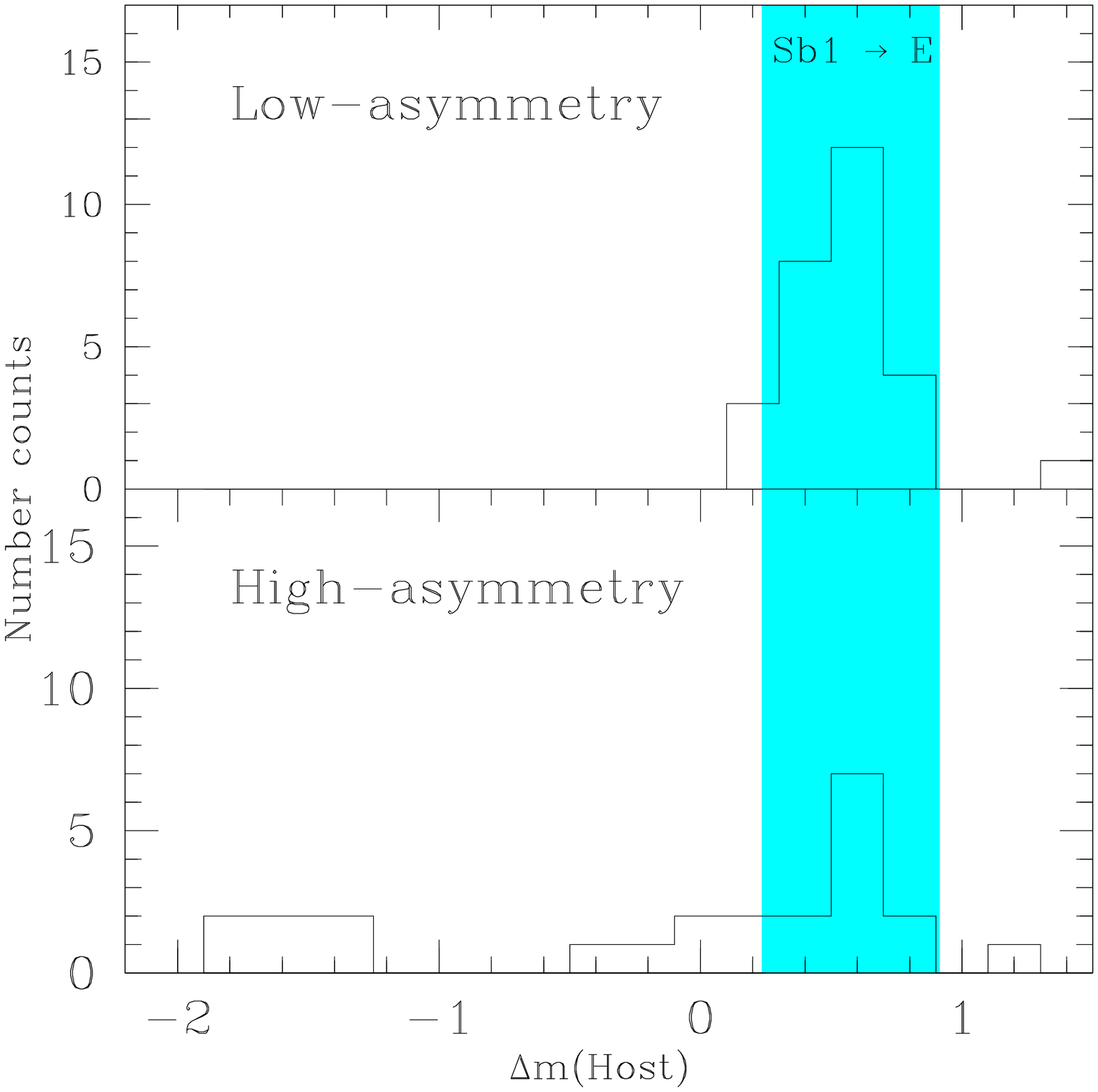}
\includegraphics[height=5.5cm]{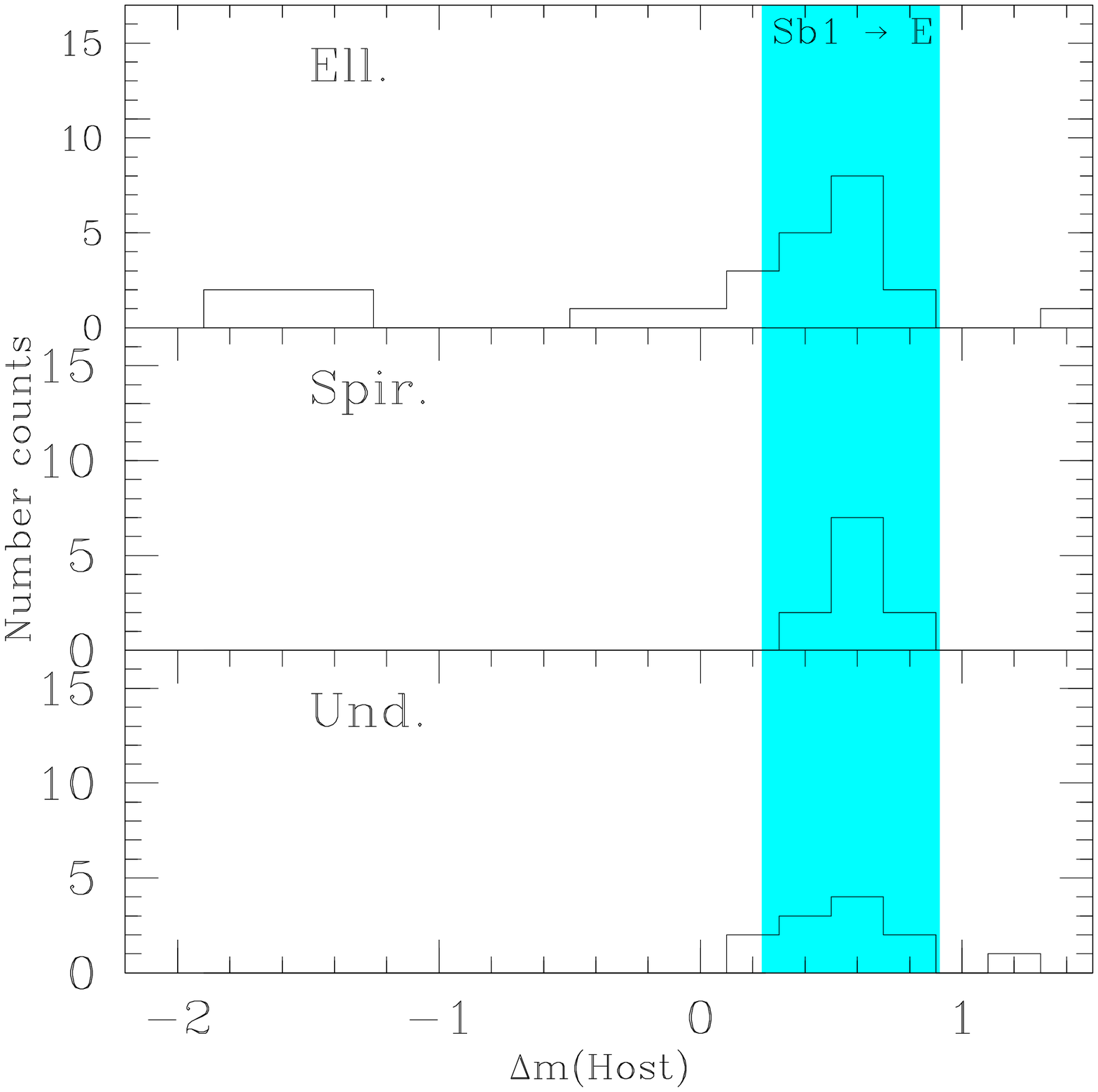}
\caption{Histogram of $\Delta m(Host)$ with $0.2$ mag bins. Typical values for template galaxies from starburst to E types are indicated by the shaded strip. Hosts with the strongest emission lines appear to the left. Left: global sample.  Middle: low and high-asymmetry subsamples, which have a different shape, the high asymmetry hosts being shifted to bluer colours, meaning that they display both stronger emission lines and/or a bluer continuum. Right:  ellipticals, spirals, and undefined morphology galaxies. Elliptical hosts have the broadest range of colours. }
\label{histo}
\end{figure*}

\subsection{Colours}
\label{coul}

As explained in Section $2$, one of the goals of the study is to compare the stellar and gas distributions in order to seek for cases where there are significant differences between both components.
We thus take the difference in apparent magnitudes between both filters, the colour $\Delta m=m_V-m_{WB}$,  for the host and the QSO separately.
As a preliminary comment, let us mention that the deconvolution process, as it treats both filters exactly in the same way for separating the QSO and host components, should have no influence on the derived colours.
Furthermore, during the deconvolution process, given the slight uncertainty on magnitudes induced by QSO-host separation as mentionned in Section \ref{mager}, some flux from the QSO might be interpreted as galaxy flux. The QSO being bluer than galaxies, it would lead to an overestimation of hosts blue colour compared to normal galaxies. Such a perturbation is however hard to estimate. It might lead to a slight shift of our hosts distribution in the histogram towards the redder galaxies in both filter. It will nevertheless not change the conclusions obtained from the colour analysis.

First of all, one might expect QSOs to be brighter in the V-filter, as it contains the strong  \hb\ and \ot emission lines. We find a mean colour for the QSOs of $\Delta m(QSO)=0.03\pm 0.17$. In order to check if this is not due to any processing error, the spectrum of a typical QSO HE1302-1017 \citep{Letawe} ($z=0.278$) has been blueshifted to $z=0.16$ and integrated in both filters. Converting the flux ratios into magnitude differences leads $\Delta m(\textrm{HE1302-1017})=0.08$, which lies well within our error bars. 

For the host galaxies, we find an average $\Delta m$(host) = 0.41$\pm 0.58$.  The left panel of Fig.~\ref{histo} shows the distribution of $\Delta m$ for all our host galaxies.  Some of these host galaxies have very blue colours $\Delta m < 0$, even bluer than the QSOs.  To identify the origins of the blue light, we compare the colours of the sample with the colours of galaxy template spectra, estimated as for QSO colours in integrating spectra in both filters. We used  galaxies of types E to Sc from \citet{Mannucci}, and from \citet{Kinney} which constructed $6$ starburst galaxies with different extinction values (from SB1 to SB6 with increasing extinction).  The associated colours are reported in Table \ref{cols} together with the colours for our sample.

\begin{table}
\centering
\begin{tabular}{cccccl}
\hline
 Galaxy&$\Delta m $ &$\Delta m $& Diff.&& $\Delta m $\\
&Global&Continuum&&&Sample\\
\hline
 E& 0.87&0.87&0&&Ellipticals\\
 S0&0.86&0.86&0&&0.27$\pm$0.78\\
 Sa&0.85&0.85&0&&\\
 Sb&0.76&0.77&-0.1&&Spirals\\
 Sc&0.65&0.67&-0.02&&0.60$\pm$0.15\\
 SB6&0.5&0.54&-0.04&&\\
 SB1&0.23&0.34&-0.11&&Undefined\\
 HE1434-1600&0.27&0.63&-0.36&&0.58$\pm$0.28\\
 \hline
 \end{tabular}
\caption{Colours estimated on galaxy template spectra and on the peculiar QSO host HE1434-1600 with a low, red continuum and strong emission lines \citep{Letawe3}. Colours are given with and without the contribution of emission lines, together with the difference in colours between both (total minus continuum only).  The last column gives the average values and scatter for our QSO host galaxies as a function of morphology}
\label{cols}
\end{table}
%We find expected colours $\Delta m(\rm{Sb1})=0.23$, $\Delta m(\rm{Sb6})=0.50$, $\Delta m(\rm{Sc})=0.65$, $\Delta m(\rm{Sb})=0.76$, $\Delta m(\rm{Sa})=0.85$, $\Delta m(\rm{S0})=0.86$ and $\Delta m(\rm{E})=0.87$, whereas our sample has a mean $\Delta m(Host)=0.41\pm 0.58$, while $\Delta m(Spir)=0.60\pm0.15$, $\Delta m(Ell)=0.27\pm0.78$ and $\Delta m(Undef)=0.58\pm0.28$. 

The colour is directly influenced by the strength of the ionization lines, but also by the slope of the continuum. First, we estimate an upper limit on the colour variation due to the continuum contribution by removing the ionization lines from the templates and compute $\Delta m$ again. The values we obtained are presented in the third column of Table \ref{cols}. It shows that the colour associated to the bluest slope for the continuum is estimated at 0.34. Moreover the slope of the continuum may be responsible for a variation in $\Delta m$ which should not go beyond $0.5$, which is the difference between the reddest (E type) and bluest (SB1 type) spectra. 
On the other hand, placing upper limits on the emission lines contribution is harder, as QSO hosts might have stronger lines than the templates spectra. The host of HE1434-1600 is a good example of this as it has particularly strong emission lines with a relatively weak and red continuum \citep{Letawe3}. From the spectrum of this QSO, we compute $\Delta m(Total)=0.27$ and, after removal of the ionization lines, $\Delta m(Continuum)=0.63$. So, the ionization lines may induce variations in $\Delta m$ of at least $0.36$. 
In conclusion, both contributions influence the result at a relatively similar level. For colours lower thant 0.34, ionized gas must be present, and for variations from templates larger than $\sim 0.5$ a bluer than expected colour cannot be explained only by young stellar population, also requiring the presence of ionized gas.

In order to analyze the influence of morphology and asymmetry on the colour distribution, we show on Fig. \ref{histo}  histograms of the colours for different hosts subsamples (global sample and subsamples with different asymmetry or morphology, from left to right). We plot the numbers of hosts whose $\Delta m$ is contained in bins of $0.2$ mag, from $-2.$ to $1.5$, and compare them to the template galaxy colours mentioned above. 

First, Fig. \ref{histo} shows that the range spanned in $\Delta m$ for our hosts is much larger that the range spanned by the templates, only covering the $0.3-0.9$ region. Moreover, the majority of the hosts seems to have colour typical of Starburst to Sc galaxies, characterized both by a large amount of ionized gas and a blue stellar continuum.

Secondly, there is a tendency for the most asymmetric hosts to be bluer. Even if the dispersion is quite large for the high asymmetry sample, ($\Delta m(\rm{high\textrm{-}a(V)\ hosts})=0.17\pm0.81$, whereas $\Delta m(\rm{low\textrm{-}a(V)\ hosts})=0.61\pm 0.24$, leading to uncertainties on the mean values of resp. $\pm0.1$ and $\pm0.04$ given the size of these subsamples), we notice that the bluest host are highly asymmetric. These objects are interpreted as hosts involved in major mergings, having a higher proportion of ionized gas and/or a bluer continuum than non-interacting ones. A similar tendancy has been found in \citet{Letawe, LetaweY}, where links between gas ionized by the QSO radiation and gravitational interactions are observed. Thus, it is tempting to propose that our observations give a statistically more relevant proof of this kind of interrelation between QSOs and their hosts. However we cannot disentangle the different ionization processes, namely ionization by the QSO, by shocks, or by stars. Consequently, the higher degree of ionization in highly asymmetric hosts might also be interpreted as due to prominent HII regions typical of strong star formation, enhanced during the interaction process.

Third,  we point out the fact that all hosts that have $\Delta m(Host)<0$, thus the bluest ones, are ellipticals. It is a suprising result as ellipticals are often supposed to be evolved galaxies with old stellar populations and only a low proportion of gas and dust. This trend for elliptical QSO host galaxies to have bluer colours was already reported by \citet{jahnke} for similar objects,  and by \citet{Sanchez}  or \citet{jahnkeapj} for samples at higher redshift, and was explained by a younger stellar component in the host. Spectral analysis of a sample of nearby QSO hosts in \citet{Letawe} also pointed to a younger-than-average stellar population. As the colour variation due to the continuum (thus the stellar population) presented above reaches at most 0.5 mag, and as the bluest slope only leads to $\Delta m$=0.34, emission line contribution has to be present in at least a dozen of outliers having $\Delta m(Host)<0$. They likely correspond to a class similar to the already studied QSO HE1434-1600 \citep{Letawe3,LetaweY}, which has a seemingly normal elliptical host, but also filamentary structures consisting of gas ionized by the central nucleus on both sides of it. This system is in gravitational interaction with a smaller neighbouring elliptical. Indeed, if the ionization was due to stars in those elliptical outliers, it would be expected to be at least as strong in spirals, which is not the case. Thus, if an ionization source has to be favoured, QSO or shock ionization clearly seems to be preferred.

We finally see in Fig. \ref{histo} (right panel) that spirals better match with classical values (inactive galaxies) than ellipticals, the latter being responsible for the broadening of the range spanned by the whole sample. \\

This hosts colour analysis reveals the difference between QSO hosts and quiescent galaxies. QSO hosts tend to have a bluer continuum, but also to contain more ionized gas than their quiescent counterparts, and this trend is stronger in ellipticals and highly asymmetric systems. It is consistent with the previous study of \citet{scoville}, who find, by analysing CO emission of $12$ low redshift ($z<0.1$) PG QSOs, that elliptical hosts are gas-rich and therefore cannot be normal ellipticals. Our findings prove that not only elliptical hosts contain more gas, but that this gas has a significant level of ionization compared to quiescent ellipticals. However, the available data do not allow to determine unambiguously the ionization source. For instance, the ionization might be due either to the QSO radiation or to shock waves during a merger event.

\subsection{Magnitude-Asymmetry relations}

\begin{figure}
\centering
 \includegraphics[width=6.5cm]{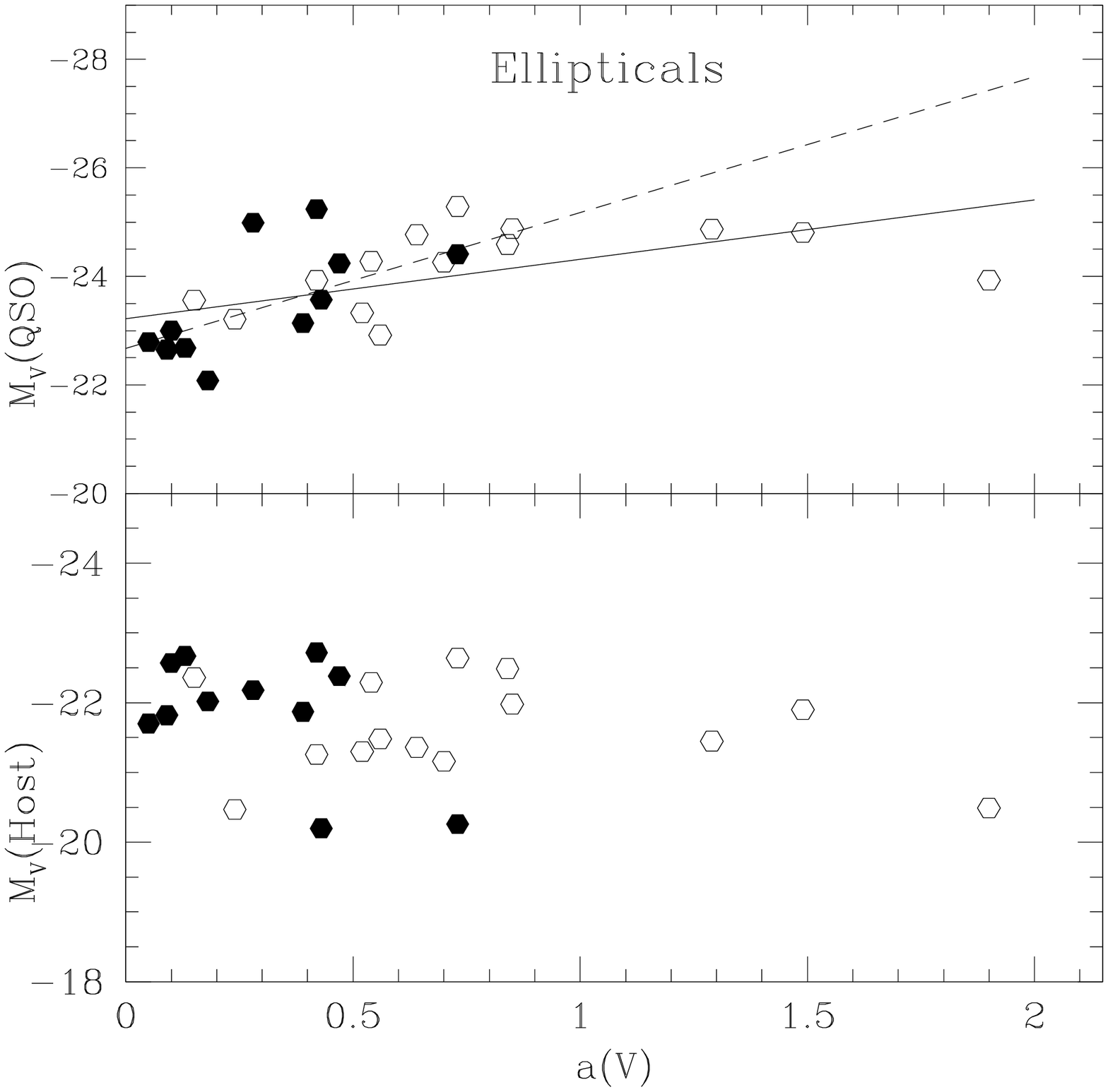}
\includegraphics[width=6.5cm]{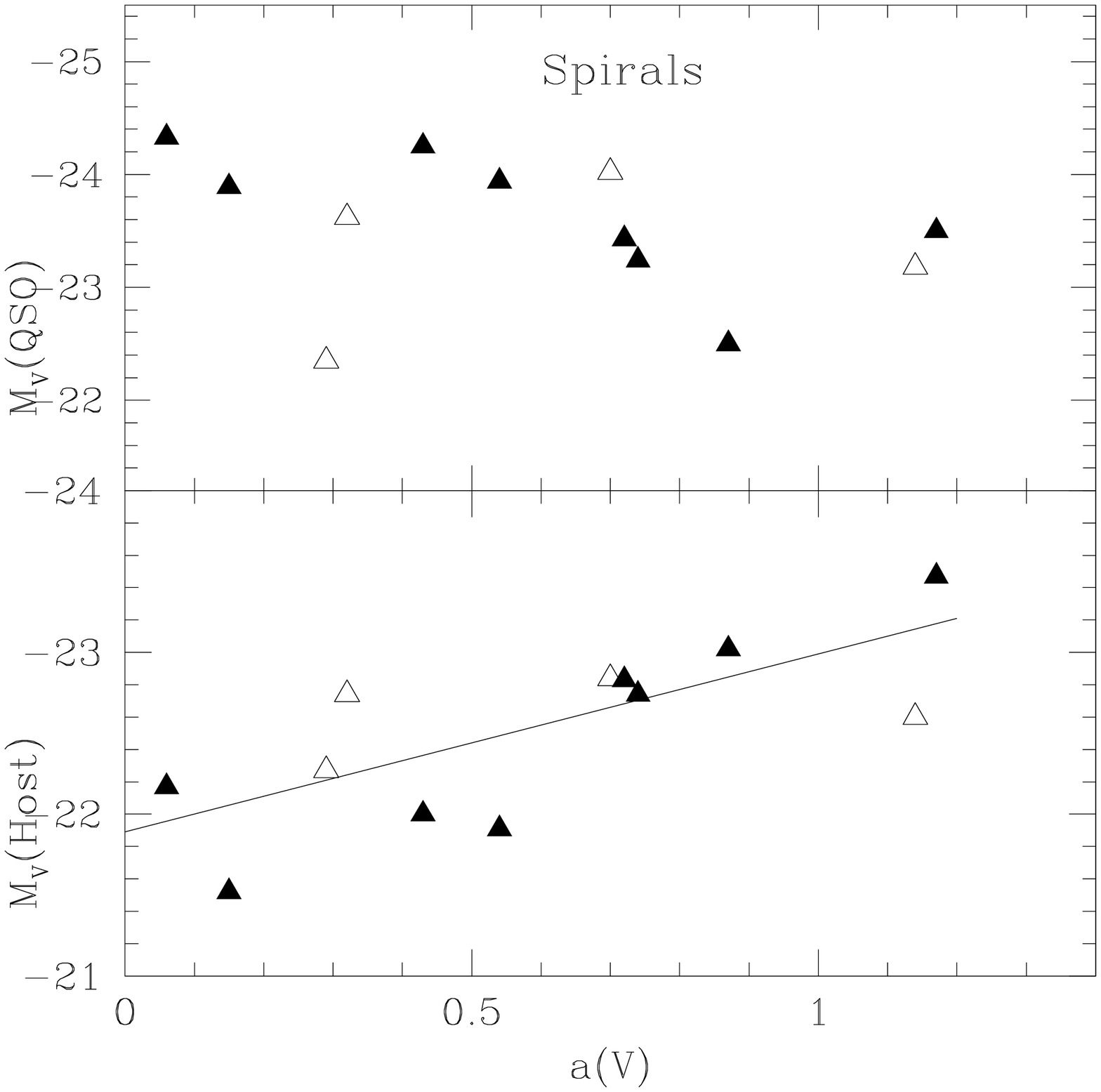}
\caption{Asymmetry is plotted against QSO and host absolute magnitude in the V filter for ellipticals and spirals separately. Open symbols indicate the presence of interactions or mergers.}
\label{asym_V_spir}
\end{figure}

If mergers and/or gravitational interactions are to trigger or enhance QSO activity, we might expect a correlation between the QSO magnitude and the degree of asymmetry of the system. The relation is shown in Fig. \ref{asym_V_spir} for ellipticals and spirals separately because they have a different behaviour. Indeed, for the ellipticals, a reliable correlation ($p=0.54$) exists between the absolute magnitude and the asymmetry coefficient (solid line in Fig. \ref{asym_V_spir}): $M_V(QSO)=(-1.10\pm 0.36)a(V)+(-23.22\pm 0.26).$
It is even more robust ($p=0.68$, dashed line) if the three higher asymmetry systems are removed from the fit, with a relation 
\begin{equation}
 M_V(QSO)=(-2.51\pm 0.61)a(V)+(-22.67\pm 0.30). 
\end{equation}\label{mVa}
A similar relation is not found for the host magnitude. 
Conversely, the asymmetry of spiral hosts does not correlate with QSO magnitude, but does correlate with host magnitude as $M_V(Host)=(-1.09\pm 0.34)a(V)+(-21.89\pm0.24),$ with $p=0.71.$ 

\begin{figure}
\centering
 \includegraphics[width=8cm,height=4cm]{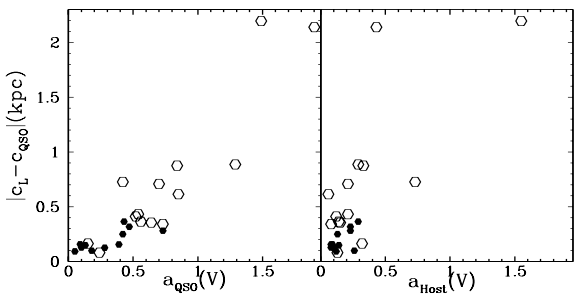}
\caption{Left (resp. right): distance in kpc between the center of luminosity of the galaxy and the QSO position against the asymmetry coefficient computed on the QSO position (resp. the center of luminosity of the host), for elliptical galaxies only.}
\label{cL_aV}
\end{figure}

For ellipticals, in the low asymmetry range ($a(V)<0.7$), asymmetry is mainly due to shifts between the position of the QSO ($c_{\rm{QSO} }$) and the center of luminosity of the host ($c_{\rm{L}}$). To illustrate this statement, we plot on Fig. \ref{cL_aV} the asymmetry according to $c_{\rm{QSO}}$ ($a_{\rm{QSO}}(V)$, left plot) and the asymmetry according to $c_{\rm{L}}$ ($a_{\rm{Host}}(V)$, right plot) versus the distance between both centers $c_{\rm{QSO}}$ and $c_{\rm{L}}$. While the distance is strongly correlated to $a_{\rm{QSO}}(V)$, it is not so much the case for $a_{\rm{Host}}(V)$, proving that, for elliptical hosts, the asymmetry is due to relatively symmetrical galaxies which are not centered on the nucleus position. In the framework of galaxy evolution via mergers, it is expected that ellipticals are created during major mergers involving galaxies of similar masses. Such mergers are also likely to trigger the activity of the nucleus. 
%However, recent findings \citep{Lobanov} prove that the proximity of two similar supermassive black holes (SMBH) can suppress the accretion disc activity via the strong tidal shears between each other. 
Thus, as the merger of the galaxies is faster than the merger of the black holes first orbiting around each other, it would not be surprizing
% if the activity is triggered via major mergers in ellipticals,
 to detect off-centered activity in apparently relaxed galaxies, where the two SBMH are still sufficiently far from each other. Alternatively, an off-centered activity might also be due to a black hole recoil with its accretion disc after the fusion of a black holes binary. Simulations from \citet{Volonteri2008} predict the existence and observability of such off-centered activity.

For spirals, the observed correlation with $a(V)$ concerns $M_V(Host)$ instead of $M_V(QSO)$. Tidal disturbances in spirals are believed to arise either from minor mergers between different mass galaxies ($4:1$ to $10:1$ or higher mass ratios, see \citet{Bournaud2004}) or in young mergers which have not evolved during enough time to reach their final state. The lack of correlation found between asymmetry and QSO magnitude supports the idea that young or minor mergers have not influenced the activity of the central nucleus yet. This idea matches well with the concluding remark in \citet{Reichard}, who suggest that the period of black hole growth may be preferencially associated with the end stages of minor mergers. 
On the other hand, morphological disturbances are already visible and contribute to raise the asymmetry in proportion to their luminosity, giving rise to the observed correlation.

\section{Conclusions}

We have studied a sample of $69$ QSOs, amongst which $60$ have been observed with the ESO NTT and SUSI2. Our QSO-host separation technique, based on the MCS deconvolution method, proves well adapted for that kind of data. \\
A morphology classification method based on the asymmetry of the system has enabled us to compare the properties of mergers and non interacting hosts. More precisely, high asymmetry systems tend to have a higher degree of ionization and a bluer continuum than low asymmetry systems.\\ 
Another interesting finding is that the QSO hosts tend to contain on average more ionized gas than quiescent galaxies. This trend is especially strong for ellipticals. It is consistent with the scenario in which AGN host galaxies lie in a transition phase in between the red and blue modes in the colour-magnitude diagram \citep{Martin}, as these emissions will make the elliptical QSO hosts appear bluer than their quiescent counterparts.  Nevertheless, our data do not allow to firmly determine if the ionization source is the QSO itself or shocks induced during merger events. \\
The relation between the QSO and host magnitudes is quite weak, and mainly found in ellipticals. The correlation observed is probably only a remnant of the more fundamental black hole-spheroid relation. We also find that high asymmetry systems cover a wider range of host magnitudes at a given QSO magnitude, which is interpreted either as luminosity excess due to galactic fusion or to particularly powerful nuclear activity.

 At this stage, we cannot firmly determine unequivocally which process is responsible for QSO triggering. Our study enlightens links between interaction processes, QSO activity and gas ionization, but it would be risky to assess unequivocally any hierarchical structure between those process. Indeed, our study does not allow to detect explicitely gas infalling to the central regions due to the interaction process. However, we find that the nuclear activity during a merging process is different for ellipticals and spirals, reinforcing the current belief that they correspond to different stages of evolution in a QSO lifetime, which fits a merger-driven evolution scheme. \\
A few ellipticals have an off-centered activity which might be due to a black hole recoil or to a merger induced activity between similar mass galaxies.
Moreover, some particular cases are reported for the first time and described in more detail in the Appendix. They seem to have experienced a wide variety of merging scenarios, and certainly deserve further investigations. Finally, we report the first image of the underlying host of PDS456, and discuss the possibility that this system may contain a double QSO.

As future works, the study of a sample of inactive galaxies in similar redshift and brightness ranges with similar spatial resolution and filters will help clarifying several open questions, such as the frequency of mergers or the spread in colours compared to templates.  A detailed comparison of the newly introduced asymmetry coefficient to other existing asymmetry measurements is also planned.

\section*{Acknowledgments}

This work was supported by PRODEX Experiment Agreement 90312 (ESA and PPS Science Policy, Belgium).

\bsp
\appendix

\section{Peculiar cases}

One of the goals of the present study is to find candidate hosts which would represent a particular stage of evolution in a QSO's lifetime (for example, similar to the now famous HE$0450$-$2958$ case studied in \citet{Mag2,LetaweY,Let4,Elbaz,Jahnke2009}), in preparation for higher resolution observations which would go deeper into the central regions and would allow to better assess the nature of QSO-host interactions. 
In the present Appendix, we review some special cases, which are essentially outliers from the main trends described in the previous section. Note that all images have a zero position angle (North is up, and East is to the right), and all indicated distances are projected distances.

\subsection{1307+085}

\begin{figure}
 \centering
\includegraphics[width=8.5cm]{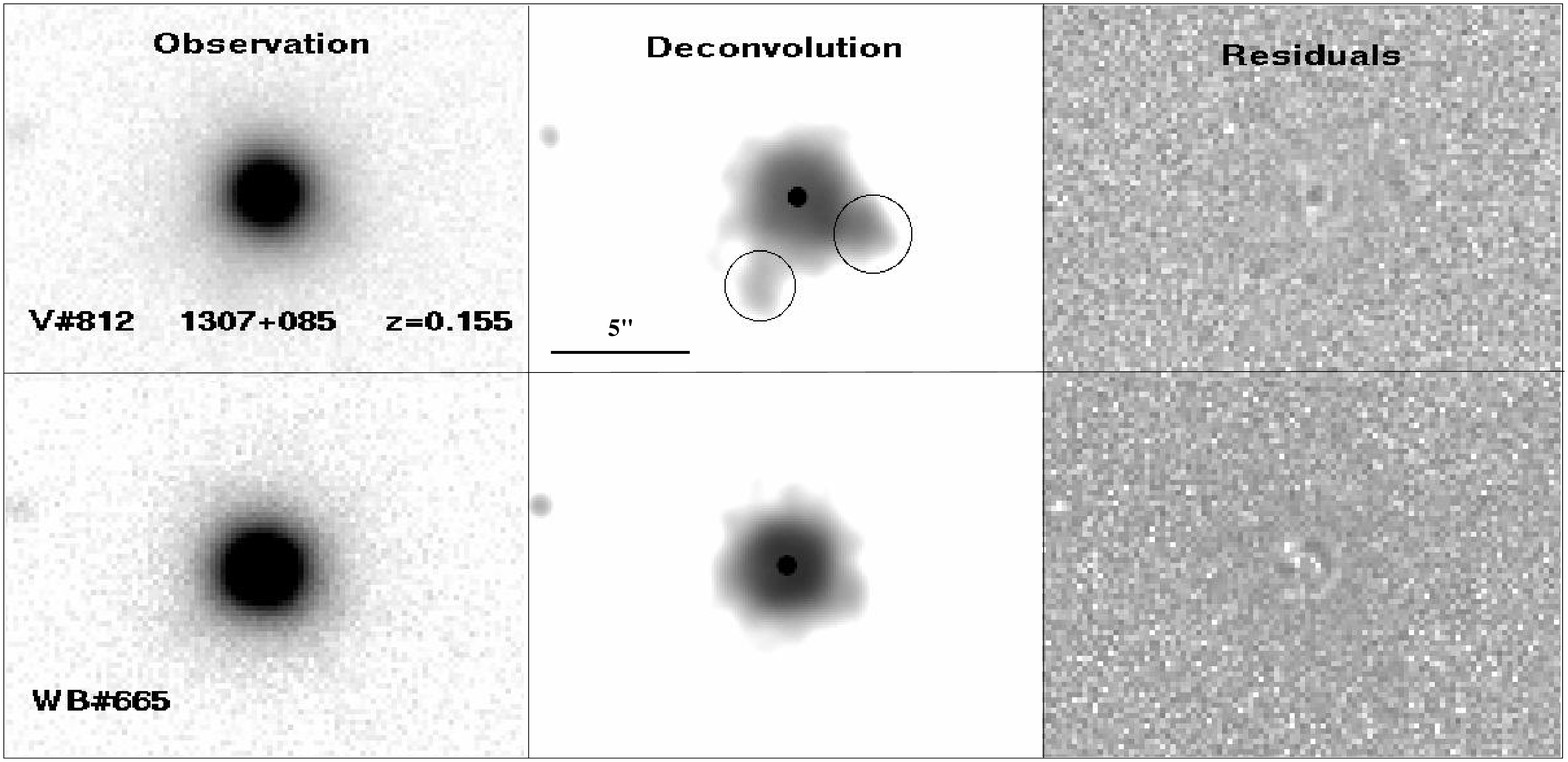}
\caption{Observations of QSO $1307+085$ in both filters. The top row (resp. bottom) shows, from left to right, an observation, the deconvolved image (with the special features discussed encircled), and the residual map for the V$\#812$ filter (resp. WB$\#665$).}
\label{1307}
\end{figure}

First of all, the host of this QSO lies amongst the very few systems which have substantially different asymmetry coefficients in the two filters ($a(V)=1.29$, $a(WB)=0.28$). Secondly, the very blue QSO colour $\Delta m(QSO)=-0.5$ suggests that the nucleus has strong emission lines compared to the continuum. Moreover, the comparison of the deconvolved images in both filters (Fig. \ref{1307}) reveals two zones (encircled in the figure) in the host that show a prominent emission in the V-filter not seen in the continuum, which we thus associate to strong emission lines. Those elements put the host in the subclass of asymmetric ellipticals with substantial gas ionization, where the strong emission zones increase the asymmetry coefficient and the magnitude in the V-filter, whereas the stellar component looks like a rather regular elliptical. 
An emission zone appears in both filters at $\simeq 22.5$kpc West from the nucleus, close to the left edge of Fig.~\ref{1307}. However we see no obvious trace of interaction between this feature and the host. 

\subsection{Q0022-2044}

\begin{figure}
 \centering
\includegraphics[width=8.5cm]{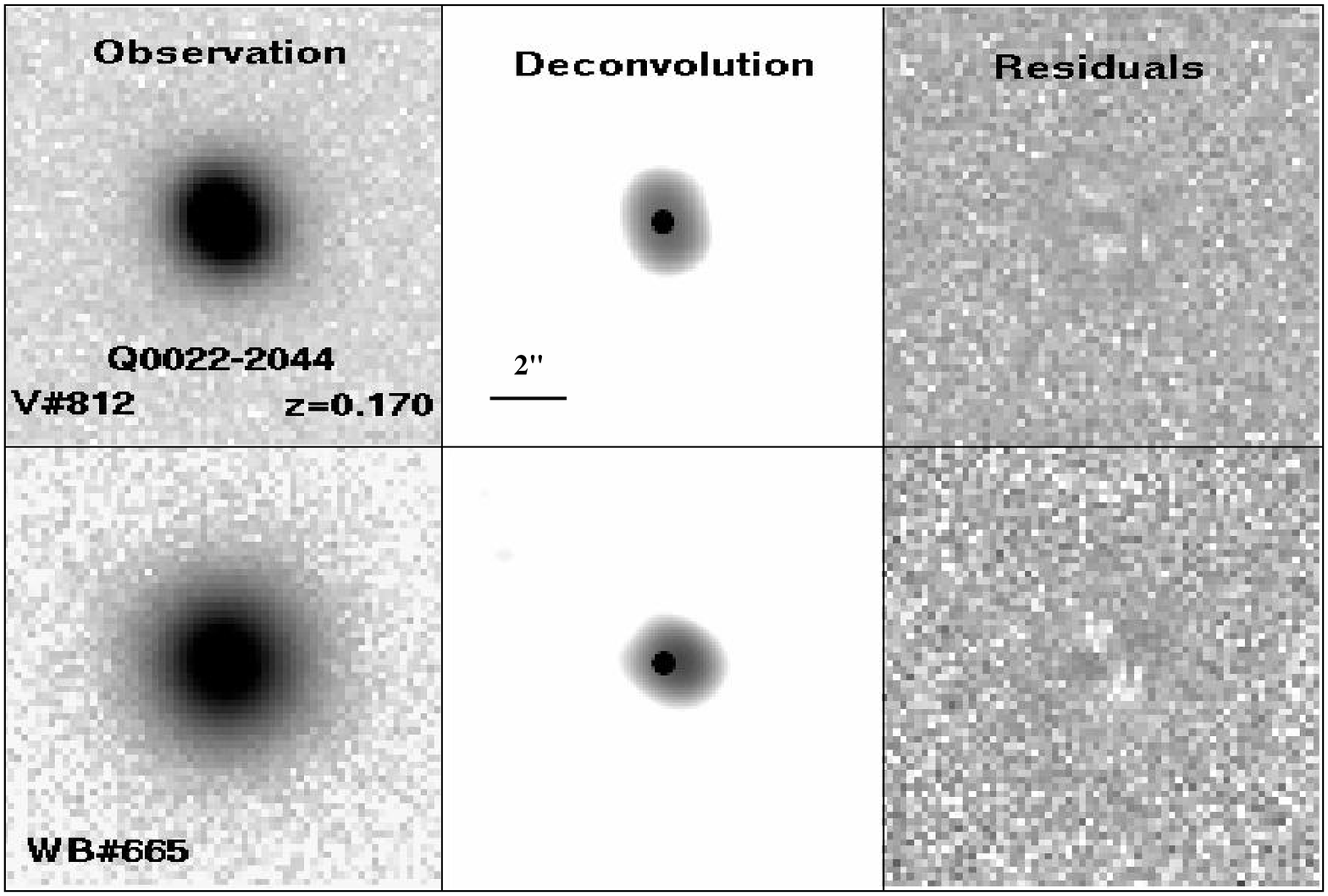}
\caption{Observations of QSO $Q0022-2044$ in both filters. The top row (resp. bottom) shows, from left to right, an observation, the deconvolved image, and the residual map for the V$\#812$ filter (resp. WB$\#665$)}
\label{Q0022}
\end{figure}

Q0022-2040 has also substantially different asymmetries in both filters ($a(V)=0.73$, $a(WB)=1.45$). The high asymmetry in the WB$\#665$ filter is due to a shift of 2 $\rm{pixels}=0.9$ kpc between the center of luminosity of the galaxy and the position of the nucleus, as shown in Fig. \ref{Q0022}. This shift, along with the very blue host colour $\Delta m(Host)=-1.94$, indicate that Q0022-2044 is a typical example of the class of ellipticals with strong gas emission lines harbouring off-centered activity already discussed in Section 5.3. Let us also mention that the QSO is particularly bright in the WB$\#665$ filter ($\Delta m(QSO)=0.41$), suggesting a very red continuum and weak emission lines.

\subsection{PDS456}
\begin{figure}
\centering
\includegraphics[width=8.3cm]{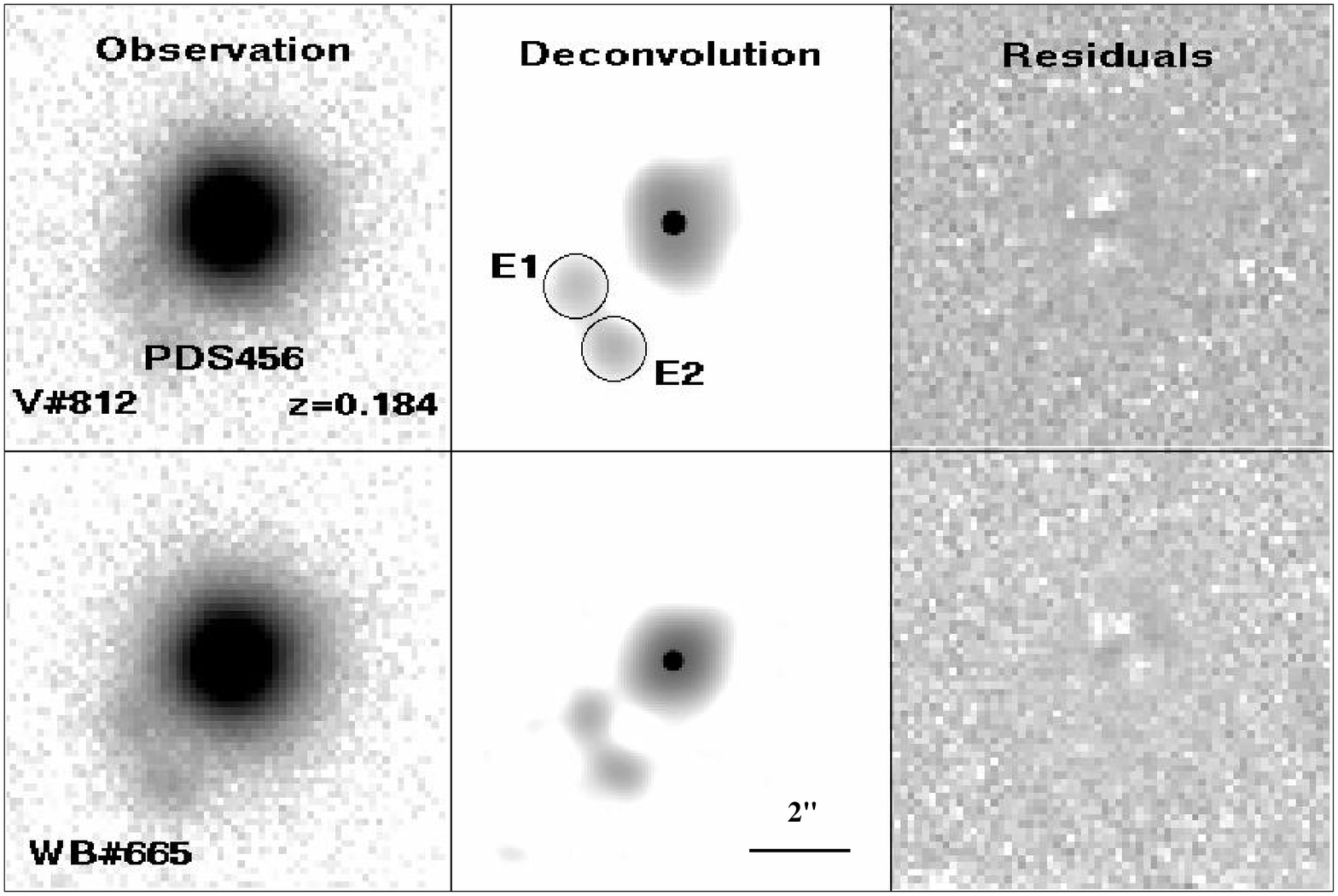}
\caption{Observations of QSO PDS456 in both filters. The top row (resp. bottom) shows, from left to right, an observation, the deconvolved image, and the residual map for the V$\#812$ filter (resp. WB$\#665$).}
\label{PDS}
\end{figure}

\begin{figure}
 \centering
\includegraphics[width=4.cm]{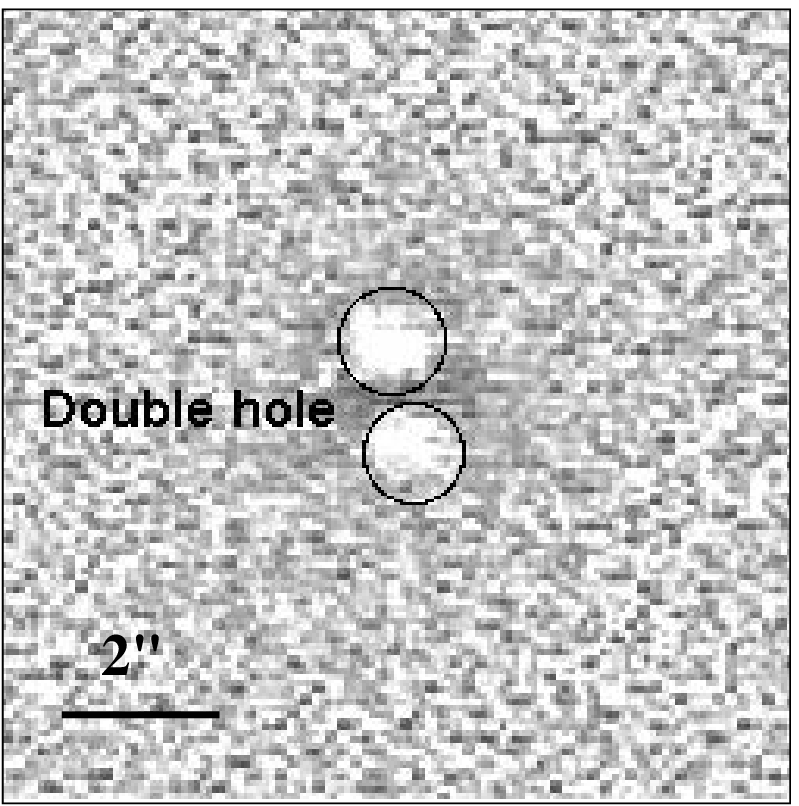}
\caption{Average residuals from the simultaneous deconvolution of $4$ observations of the QSO PDS456 in the V filter. The two holes in the residuals suggest the presence of a second point source.}
\label{PDS_resi}
\end{figure}

\begin{figure}
 \centering
\includegraphics[width=7.6cm]{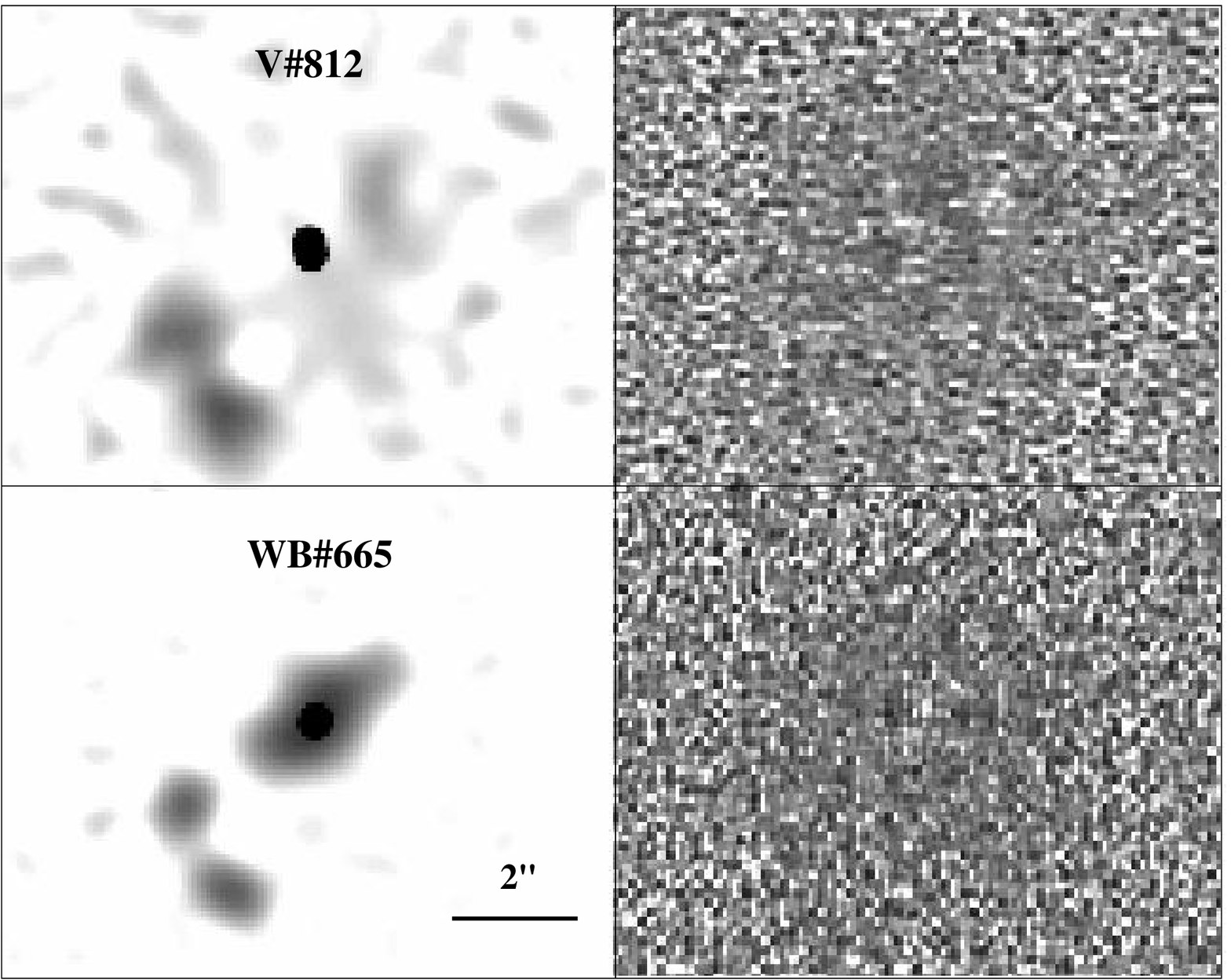}
\caption{Deconvolved image in both filters of PDS456, with two point sources very near from each other. The average residuals are considerably improved, without any significant structure remaining.}
\label{PDS_2src}
\end{figure}

PDS$456$ is known as the most powerful QSO in the local Universe ($z<0.3$). It is thought to be a radio-quiet analogue of the famous $3C273$ \citep{Schmidt}. Spectra have already been obtained in basically all wavelength ranges and K-band, radio and CO(1-0) images are also available  (see \citet{obrien} or \citet{yun} for a complete review and references). It was suggested that it is in a critical transition phase between an Ultra Luminous InfraRed Galaxy (ULIRG) and a QSO \citep{Sanders}, as it shows an optical spectrum dominated by broad emission lines, large IR and X-ray luminosities, as well as a large dust/cold gas content. Moreover, UV and X-ray spectra reveal the presence of decelerating cooling outflows probably driven by radiation or magnetic field. The deconvolved image (Fig. \ref{PDS}) shows two compact sources indicated by E1 and E2,situated a $~sim 2$ ercseconds S-W of the QSO, and compatible with the three blended sources of the K-band image in \citet{yun}. Deconvolution allows to disentangle the underlying host, which seems elliptically shaped. It is, to our knowkedge, the first image of the galaxy hosting this QSO in the visible. Magnitudes and asymmetries are given in Table \ref{mag} for the whole system (E1+E2+Host), and decomposition in the different components is given in Table \ref{PDS_2_mag}.  The difference in magnitude between the host and the QSO ($-4.14$) is the highest of the sample, which indicates a particularly strong activity of the nucleus. \\
Let us now focus on the residuals map shown in Fig. \ref{PDS} for the V filter. It has a structure typical of the presence of two blended point sources with similar intensities. Indeed, if a point source is to be fit where in fact there are two closely blended ones, the two peaks will not be correctly fit and this will result in a double hole in the residual map (model minus observation). If such a feature is real, and does not correspond to a PSF mismatch on that particular observation, we expect to find it in each exposure. Figure \ref{PDS_resi} shows an average of the four residuals resulting from the simultaneous deconvolution of 4 observations. The double-hole structure is clearly present. We tested the hypothesis of a double nucleus by adding in the deconvolution process a second point source, along with the usual diffuse background. The result is displayed on Fig. \ref{PDS_2src}. In the V filter, the two point sources, of comparable magnitude $M_V\sim-27.45$ are only separated by $0.51$kpc and the host nearly disappears, with only a faint tail starting from the nucleus. For the WB filter, the separation is only $0.22$kpc, and a faint host, whose morphology can hardly be determined because of its faintness and its small angular size, is detected. Magnitudes in both filters for the two point sources fit are given in  the bottom part of Table \ref{PDS_2_mag}.

\begin{table}
 \centering
 \begin{minipage}{140mm}
%  \caption{List of the QSOs already observed}
  \begin{tabular}{lrrrrr}
  \hline
        &  \multicolumn{2}{c}{$m(QSO)$}&  $m(Host)$ & $m(E1)$& $m(E2)$ \\
 V	&\multicolumn{2}{c}{13.42} & 17.56 & 19.35 & 19.27\\
 WB	&\multicolumn{2}{c}{13.10 }& 16.43 & 18.53&18.34\\
% \hline
      &  $m(S1)$ & $m(S2)$&  & &  \\
%\hline
V	&14.17 &14.17 & 17.56 & 19.35 & 19.27\\
 WB	&13.31 &14.99 & 17.10 & 18.73&18.54\\
 \hline
\end{tabular}
\end{minipage}
\caption{Magnitudes of the different components of the deconvolution of PDS456 with one (top) or two (bottom) point sources at the center of the host.}
\label{PDS_2_mag}
\end{table}

Let us now examine arguments in favour and against the double active nucleus hypothesis. The very good quality of the deconvolution with two point sources makes clear that there is at least a very compact source in both filters next to the ``main'' nucleus. Moreover, for the one point source fit, the value $\Delta m(Host)=1.13$ is the highest of our sample and surpasses the typical elliptical value, indicating an abnormally red continuum. This strong compact continuum is better fitted with a point source, in coherence with an hypothetical AGN activity. The presence of two separate nuclei might explain its high luminosity, and is not inconsistent with the accepted scenario of a transitory system between ULIRG and QSO via a merging process. It could also account for the exceptionally broad absorption \citep{obrien} and emission \citep{yun} features observed.  \\
Conversely, if both sources were to correspond to AGN activity, we would expect a common center in both filters, which is not really the case ($\delta c\sim 0.32$kpc). Also, their magnitudes behave differently in each filter (see Table \ref{PDS_2_mag}), their relative intensities being much more different in the WB filter than in the V filter. Moreover, the value $\Delta m(Host)=1.13$ might be interpreted as due to a strong reddening of the host, which would match the ULIRG hypothesis. Thus, from our observations, it is risky to assess their true nature unequivocally.\\
All in all, it is clear that PDS456 is an exceptional object, but maybe it is exceptional in a presently unexpected way. Deep high resolution optical imaging, or 3D integral field spectroscopy, processed with a similar method as the one used in \citet{LetaweY}, might help clarifying its status.

\subsection{PG 1012+008}

\begin{figure}
 \centering
\includegraphics[width=8.3cm]{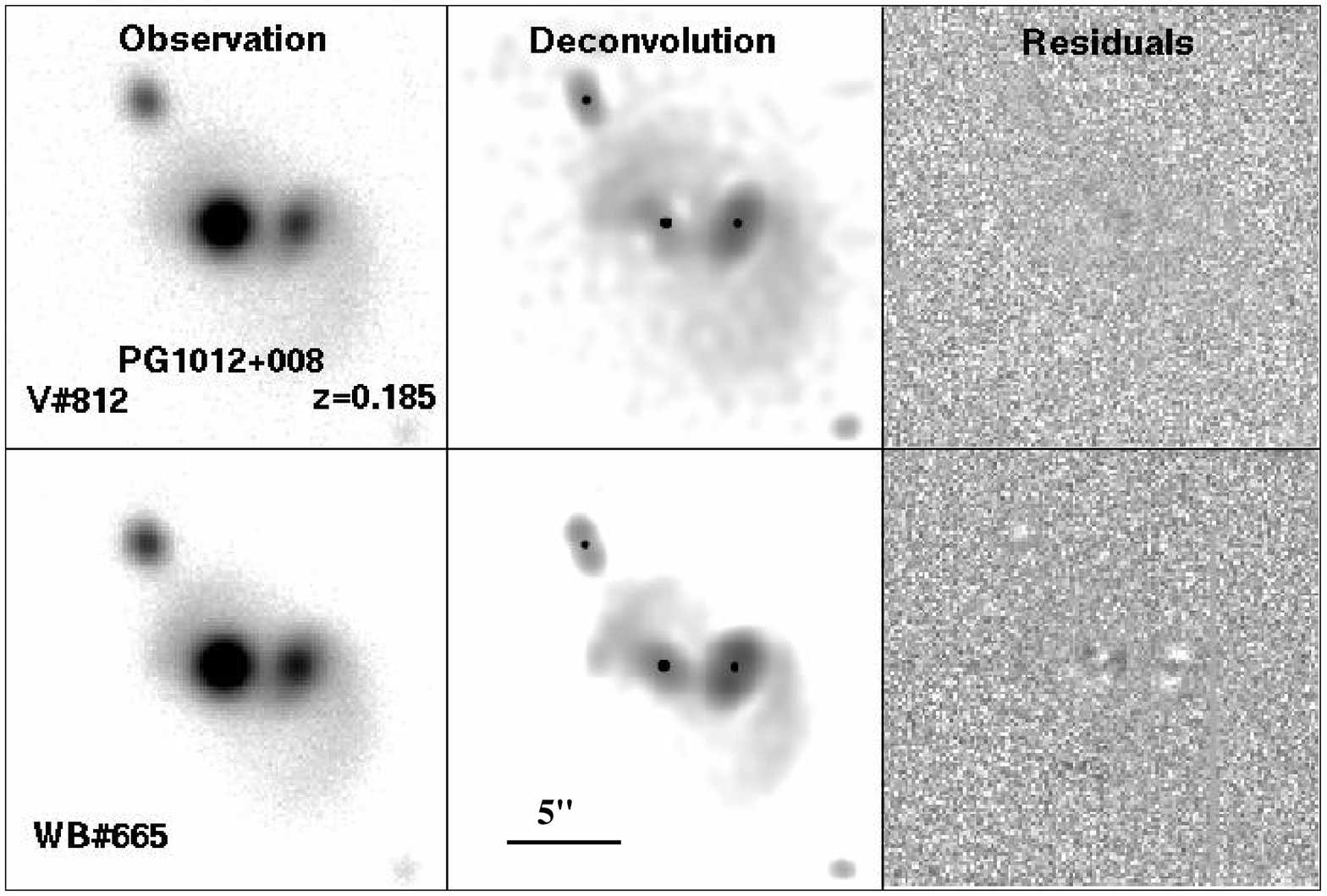}
\caption{Deconvolution process for PG$1012+008$, where $3$ point sources are included, which don't necessarily correspond to any AGN-activity.}
\label{1012}
\end{figure}

\begin{figure}
 \centering
\includegraphics[height=8.cm,angle=-90]{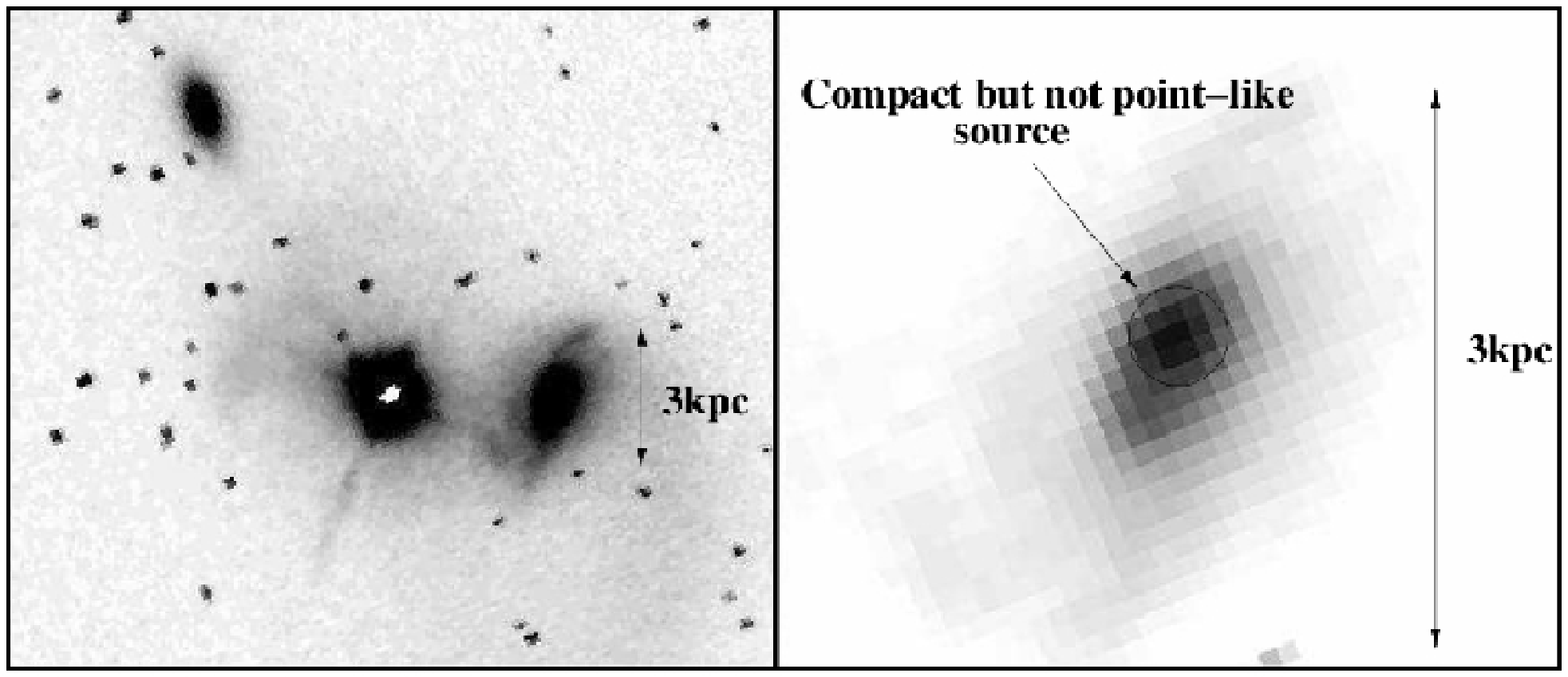}
\caption{PG$1012+008$ observed with HST/WFPC2 and a zoom on the companion galaxy containing a compact, but not point-like, emission zone.}
\label{PG1012_wfpc}
\end{figure}

From Fig. \ref{1012}, it is clear that PG$1012+008$ is involved in a merger containing at least $2$ galaxies separated by $\simeq 10$kpc, and a third one $20$kpc N-W from the QSO, which might also be interacting gravitationally. The best fit is obtained by using $3$ point sources in the deconvolution, located at each galactic center position. However, \citet{Bahcall97} have already observed this system with HST and WFPC2, and no other point source was found. This difference may be explained in the following way.  Given the difference in resolution between both observations ($0.1$''/pxl for WFPC2 compared to $0.161$''/pxl for SUSI2), a very compact and intense stellar emission just resolved by WFPC2 might look unresolved in our SUSI2 observations. At the QSO redshift, the range of sizes in which an emission zone would be resolved with WFPC2 but not with SUSI2 is $0.2-0.5$kpc. We analysed the morphology of those two putative point-like sources using the WFPC2 archives. Figure \ref{PG1012_wfpc} reveals that the point-like source nearest from the QSO looks like a compact emission zone whose FWHM is 4 about pixels. The QSO FWHM in that image being only 2 pixels, this emission is clearly not point-like. However, this 4 pxl width converts in SUSI2 pixels to 2.5 pxl FWHM, which is very close to the resolution fixed in the deconvolution process (2 pixels FWHM), making it look very much like a point-source at this resolution. Moreover, the flux ratio  between the QSO and the compact emission is the same ($\simeq0.045$) as the flux ratio between the QSO and the point source in the SUSI2 images, reinforcing the hypothesis that the compact emission in WFPC2 and the second point source in SUSI2 are the same object. The same arguments hold for the third point source.
Another indication that the additional compact regions are not related to any AGN activity is that the colours computed for the nearest galaxy center and the other one are $\Delta m =0.57$ and $0.62$. Those values are higher than for any QSO in our the whole sample, revealing an important contribution of the continuum compared to emission lines. This fact tends to favour the view that the $2$ added point sources do not correspond to AGN activity, but rather to compact luminous stellar regions, characterized by a prominent continuum. 
\subsection{1151+117}
\begin{figure}
 \centering
\includegraphics[width=8.3cm]{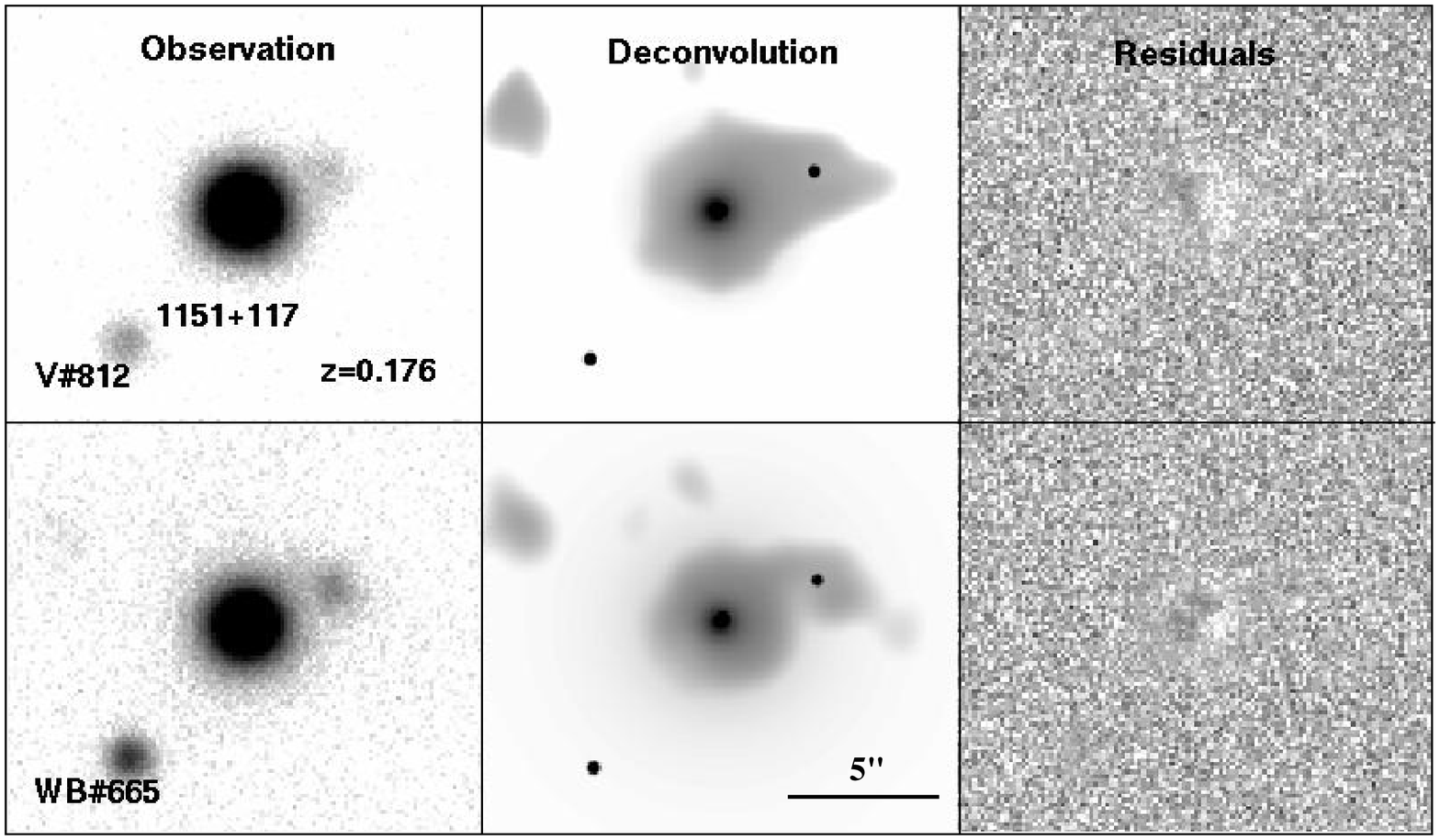}
\caption{Deconvolution of 1151+117, with $3$ point sources. The S-W one is probably a foreground star, whereas the two others are hosted by interacting galaxies.}
\label{1151}
\end{figure}

Apart from the central QSO, the observation of 1151+117 (Fig. \ref{1151})  reveals the presence of two emitting regions $20.6$kpc S-W, and $11.5$kpc N-E of the QSO. Both need a point source component to achieve good residuals. Deconvolution reveals that the S-E one is well fit by a single point source, suggesting it might be a foreground star. On the other hand, the N-E one needs a smooth background to achieve a good fit. This point source is quite weak ($M_V=-19.32$), and has a dominant continuum emission $\Delta m=0.33$. Similarly to PG1012+008, this high value of $\Delta m$ might indicate that the point-like emission is due to a compact stellar region, even if it is still within the range of reasonable value for an active nucleus. \\
The background needed around the second point source shows clear signs of gravitational interactions with the host, especially on the WB$\#665$ deconvolved image, where the galaxies seem to share a tidal tail, suggesting a configuration similar to the well known M51 \citep{Salo,Toomre}, where two galaxies of mass ratio $3:1$, viewed nearly face-on, probably encountered for the first time $300\pm100$ Myr ago. Two additional faint emissions are also present in both filters (N-W and N), and may also be linked to the merging process.  
 
\subsection{HE1202-0501}
\begin{figure}
 \centering
\includegraphics[width=8.cm]{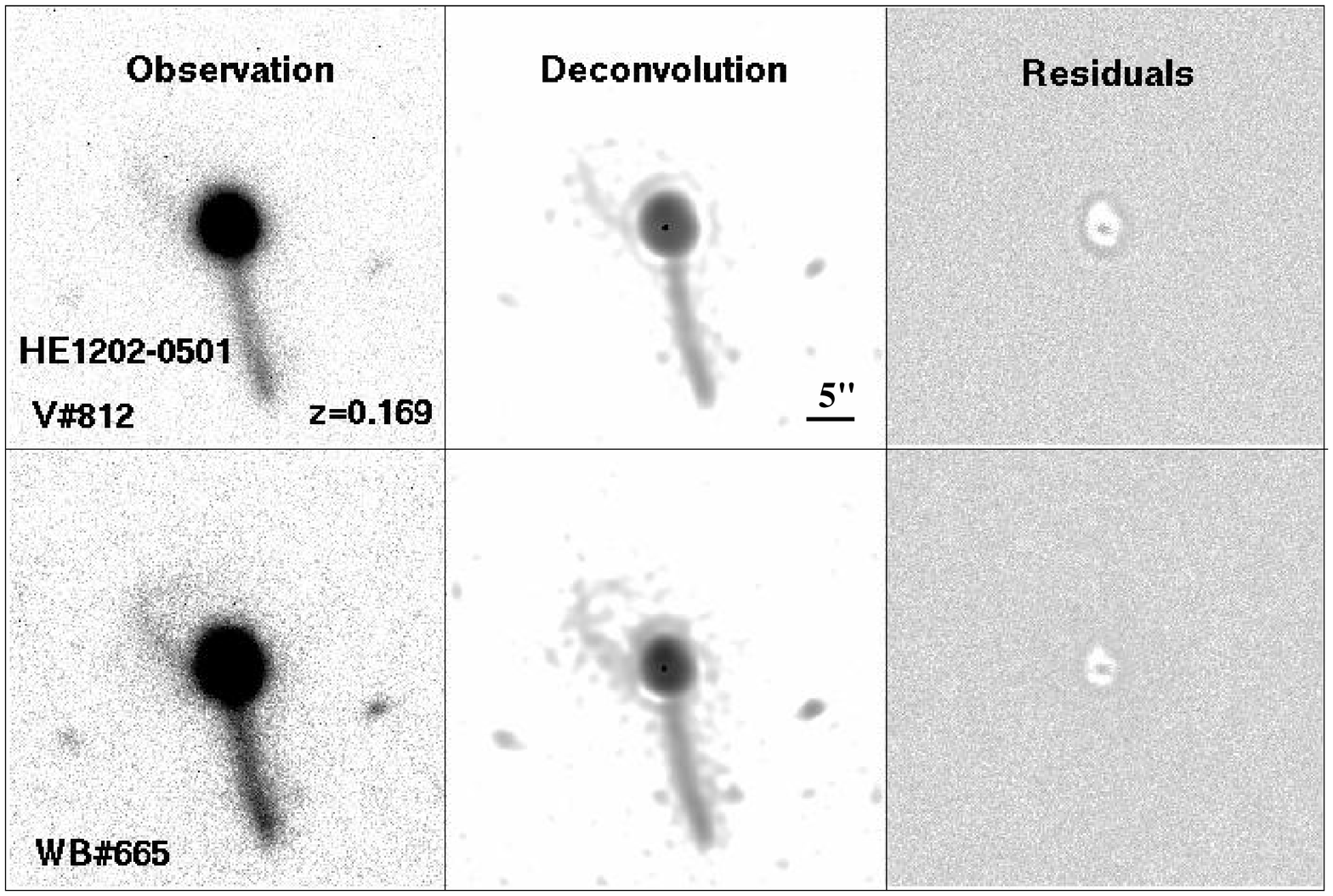}
\caption{Deconvolution process for HE$1202-0501$. The very atypical morphology may be created by the collision of two galaxies of different disc mass. The rather poor quality of the residuals is likely due to significant short-scale structures near the center.}
\label{HE1202}
\end{figure}

The host of HE1202-0501 (Fig. \ref{HE1202}) is very atypical, with a $45$ kpc nearly straight tail extending from the center towards the S, and containing a few minor substructures. Another much weaker tail of $\sim 27$ kpc length extends from the center to the N-W. Such a special configuration is very similar to the shape of a $1.2-1.3$ Gyr old major merger of nearly equal mass disc galaxies (\citet{Springel}, Fig. 7). These simulations also show very complex intense features next to the center.  If such structures are prensent in HE1202-0501, they might not be resolved by SUSI2, but could substantially contribute to the observed total central nucleus luminosity, leading to PSF mismatches and thus decreasing the quality of the deconvolution around the center, as can be seen in the residuals shown in Fig. \ref{HE1202}. 
HE$1202-0501$ also looks very similar to the famous ``Mice'' merger, NGC$4746$, which is also well modelled \citep{Toomre,Mihos,Barnes} by a two identical mass disc galaxies merger event, where one of the two discs is viewed edge-on and mimics an elongated bar. 
They are a few major differences between NGC$4746$ and HE1202-0501. In ``The Mice'', the galactic centers are separated by $20$kpc, and both tails have approximately the same length and luminosity, whereas HE1202-0501 shows only one nucleus, and different size tails. Different tail size and luminosities are suggestive of different mass galaxies. 
We thus conclude that HE1202-0501 is most probably a merger between two disc galaxies of different masses containing a sufficient amount of gas to induce AGN activity, one of the two discs being viewed edge-on. The poor residuals near the center may result from intense star formation activity around the nucleus.

\subsection{1023-014}

\begin{figure}
 \centering
\includegraphics[width=8.cm]{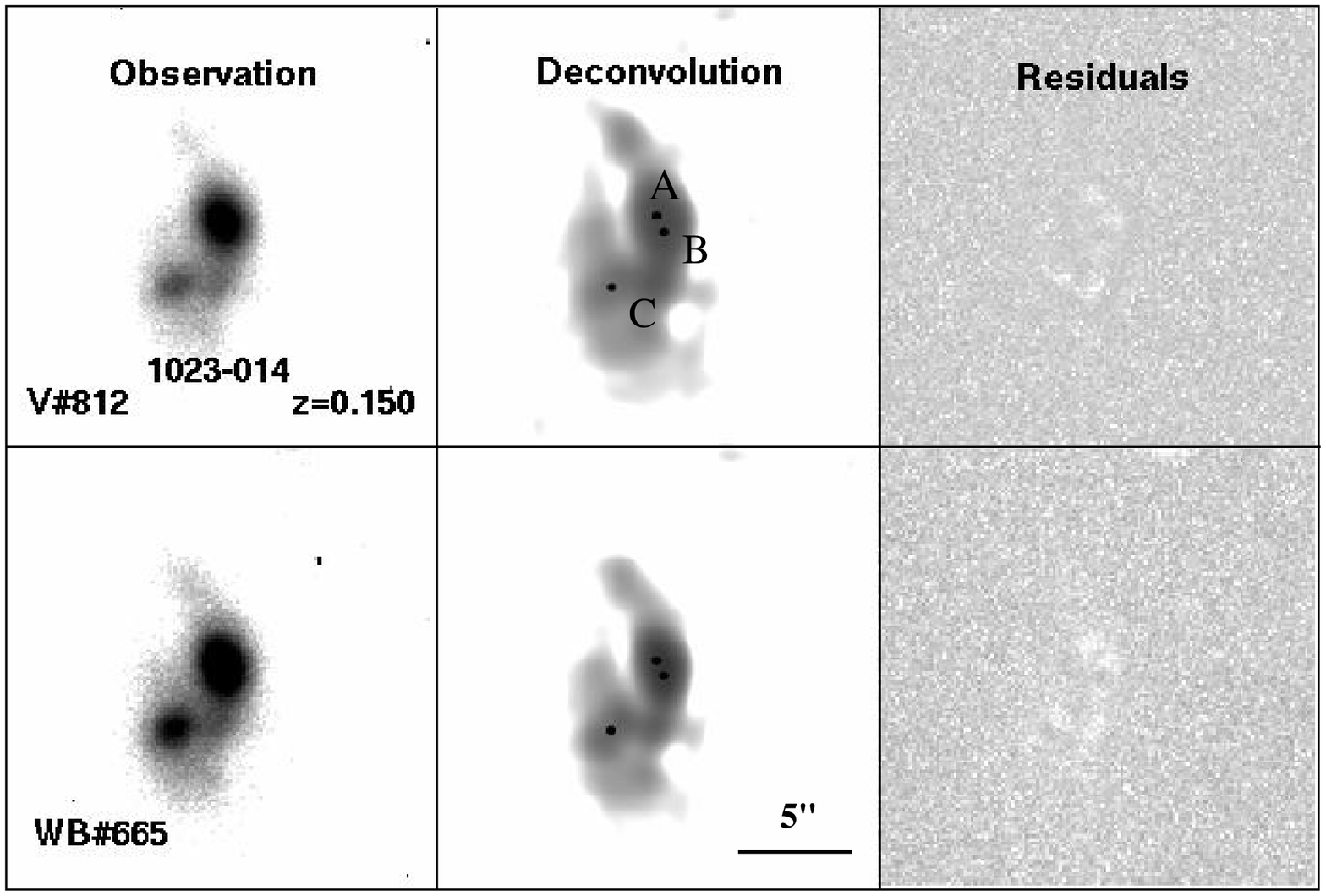}
\caption{Deconvolution process for the major merger of the QSO $1023-014$. $3$ point sources are necessary to achieve satisfactory residuals.}
\label{1023}
\end{figure}

The QSO $1023-014$ is involved in a major merger involving two galaxies of comparable size. The best deconvolution is achieved with $3$ point sources, labelled A, B and C on Fig. \ref{1023}. While A and B have are very similar in both filters ($m_V(A)=19.54,$ $m_V(B)=19.67$, $\Delta m(A)=-0.06$, $\Delta m(B)=-0.01$), C has a quite different behaviour, with $m_V(C)=20.55$ and $\Delta m(C)=0.51$, which deviates strongly from the mean $\Delta m(QSO)$. As in the case of PG $1012+008$, it indicates that the source C might not be related to AGN activity but rather to strong unresolved stellar emission. Moreover, this system lies significantly below the $M_V(QSO)\textrm{-}M_V(Host)$ relation, having an underluminous AGN compared to its host. $M_V(QSO)-M_V(Host)=1.27$ is the highest value of the whole sample, the nuclear magnitude being indeed below the standard delimitation between QSO and Seyfert. All in all, and given the fact that the QSO and total host are more powerful in the WB$\#655$ filter (the one containing no emission line), $1023-014$ could be a rather dry merger, with no sufficient amount of gas to create powerful AGN activity, but possibly one or two low luminosity AGNs. Further studies, including spectroscopy of the nucleus, would be necessary to test this hypothesis.  

\subsection{HE1211-1905}

\begin{figure}
 \centering
\includegraphics[width=8.cm]{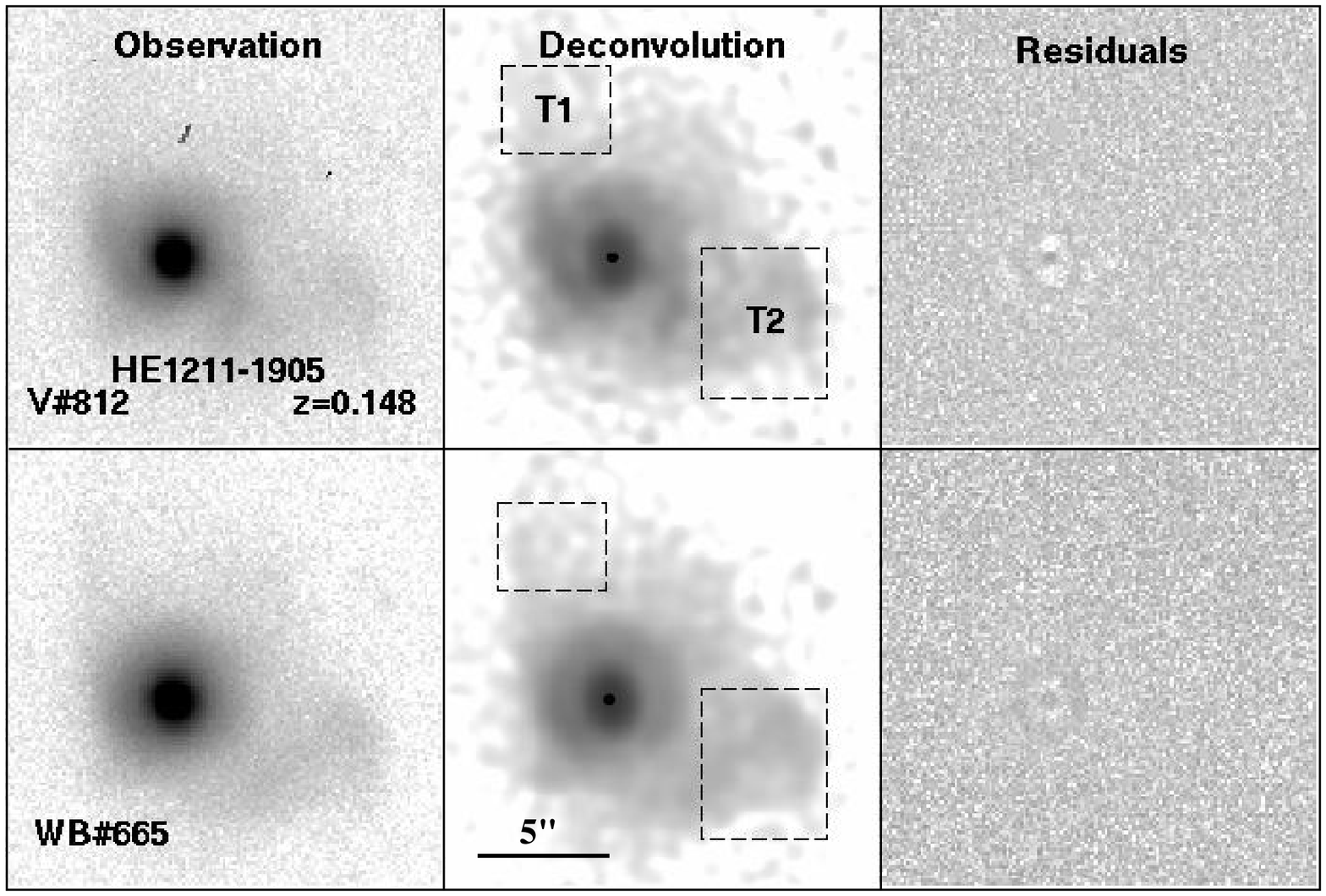}
\caption{Deconvolution process for the strongly disturbed host of the QSO HE1211-1905.}
\label{HE1211}
\end{figure}

\begin{figure}
 \centering
\includegraphics[width=8.4cm]{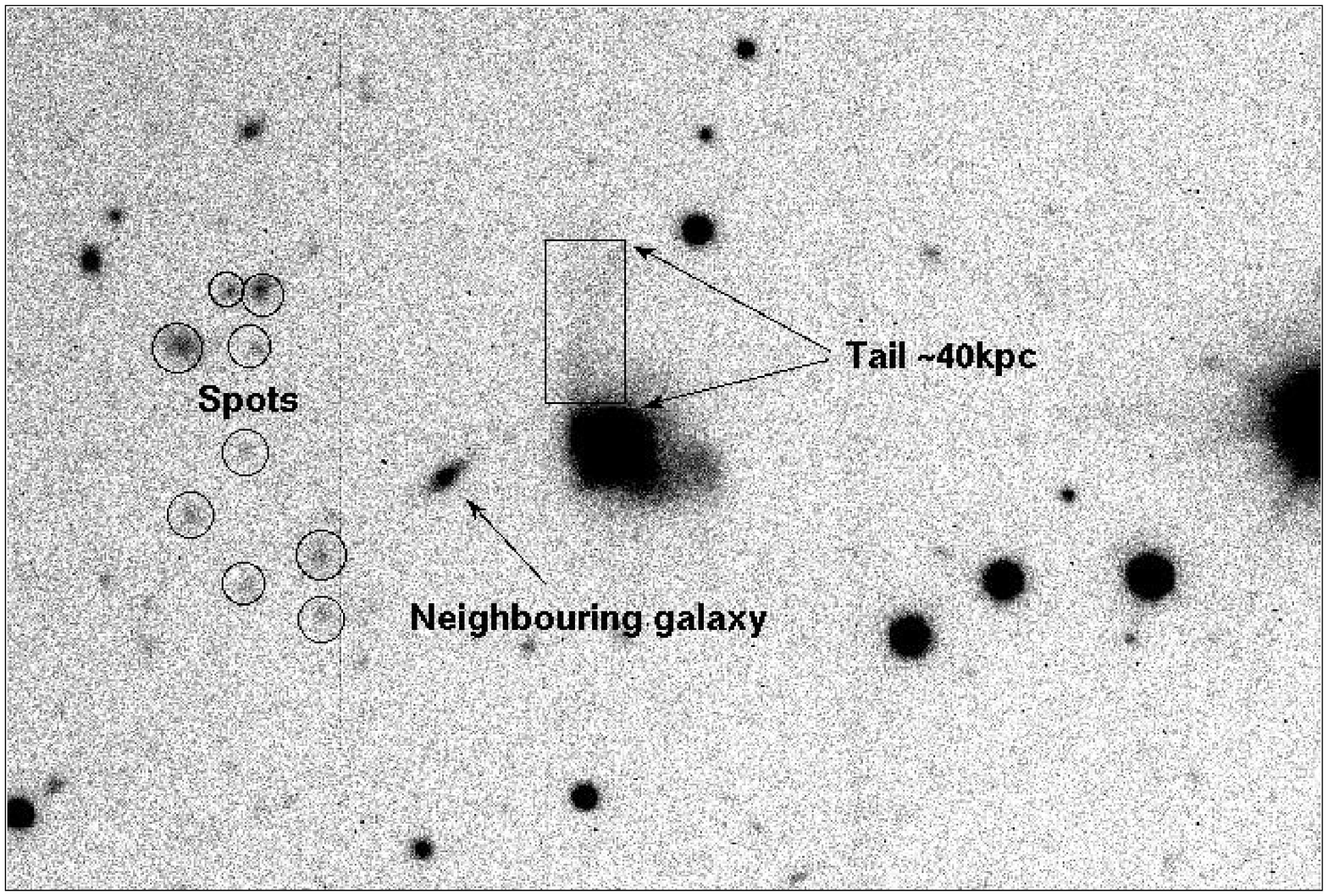}
\caption{Surrounding of HE$1211-1905$. The tidal tail of $\sim 40$kpc is surprisingly long. A companion galaxy and a few spots which are present on the same side may be linked to the host disturbances.  }
\label{HE1211T1}
\end{figure}

The QSO HE1211-1905 is surrounded by a spiral galaxy which shows clear signs of gravitational interaction via extended tails N and E of the nucleus, corresponding to the T1 and T2 boxes in Fig. \ref{HE1211}. Given the significant area of those tails ($\sim10\rm{ kpc}^2$), it makes sense to check differences in magnitudes between both filters in those regions in order to specify their nature. \\
The analysis reveals that T1 and T2 are very similar, as they both have $\Delta m=-0.05$. This negative value is much lower than the value computed for the whole host ($0.5$). It indicates that ionized gas is more prominent in the faraway regions T1 and T2 than in the host's central regions.
The tail T1 goes even beyond the extracted image. Fig. \ref{HE1211T1} shows a larger area around the QSO, where the size of T1 can be estimated to $\sim 40$ kpc. The cause for such disturbances is unclear because HE1211-1905 appears quite lonely, appart from a tiny galaxy, probably elliptical, lying $35$ kpc from it. Let us mention that there are also a few spots in the overall direction of that companion galaxy, which may also be linked to the system. Another possibility is that the system is a late-stage merger, where the two galaxies nuclei have rapidly merged, leaving a host not yet relaxed.

\label{lastpage}

\end{document}